\documentclass[journal=jpcbfk,manuscript=article,layout=twocolumn]{achemso}
\SectionNumbersOn
\usepackage[version=3]{mhchem} 
\usepackage{bm, amssymb, color}
\usepackage[caption=false]{subfig}
\usepackage{ulem}

\newcommand{\fn}[2]{\mathinner{#1\mathopen{\left(#2\right)}}}
\newcommand{\vect}[1]{\bm{#1}}
\newcommand{\spD}[1]{\fn{\tilde{\chi}_{_V}}{#1}}

\author{Charles Emmett Maher}
\affiliation{Department of Chemistry, Princeton University, Princeton, New Jersey 08544, United States}
\author{Frank H. Stillinger}
\affiliation{Department of Chemistry, Princeton University, Princeton, New Jersey 08544, United States}
\author{Salvatore Torquato}%
\email{torquato@princeton.edu}
\affiliation{Department of Chemistry, Princeton University, Princeton, New Jersey 08544, United States}
\alsoaffiliation{Department of Physics, Princeton University, Princeton, New Jersey 08544, United States}
\alsoaffiliation{Princeton Institute for the Science and Technology of Materials, Princeton University, Princeton, New Jersey 08544, United States}
\alsoaffiliation{Program in Applied and Computational Mathematics, Princeton University, Princeton, New Jersey 08544, United States}

\title[An \textsf{achemso} demo]
  {Kinetic Frustration Effects on Dense Two-Dimensional Packings of Convex Particles and Their Structural Characteristics}

\begin{document}



\begin{abstract}
The study of hard-particle packings is of fundamental importance in physics, chemistry, cell biology, and discrete geometry. 
Much of the previous work on hard-particle packings concerns their densest possible arrangements.
By contrast, we examine kinetic effects inevitably present in both numerical and experimental packing protocols.
Specifically, we determine how changing the compression/shear rate of a two-dimensional packing of noncircular particles causes it to deviate from its densest possible configuration, which is always periodic.
The adaptive shrinking cell (ASC) optimization scheme maximizes the packing fraction of a hard-particle packing by first applying random translations and rotations to the particles and then isotropically compressing and shearing the simulation box repeatedly until a possibly jammed state is reached.
We use a stochastic implementation of the ASC optimization scheme to mimic different effective time scales by varying the number of particle moves between compressions/shears. 
We generate dense, effectively jammed, monodisperse, two-dimensional packings of obtuse scalene triangle, rhombus, curved triangle, lens, and ``ice cream cone" (a semicircle grafted onto an isosceles triangle) shaped particles, with a wide range of packing fractions and degrees of order.
To quantify these kinetic effects, we introduce the kinetic frustration index $K$, which measures the deviation of a packing from its maximum possible packing fraction.
To investigate how kinetics affect short- and long-range ordering in these packings, we compute their spectral densities $\spD{\vect{k}}$ and characterize their contact networks. 
We find that kinetic effects are most significant when the particles have greater asphericity, less curvature, and less rotational symmetry. 
This work may be relevant to the design of laboratory packing protocols.
\end{abstract}

\section{Introduction}
Dense hard-particle packings have been employed to model the behavior of simple liquids, glasses, and crystalline states of matter \cite{hughel_1965, zallen_2007, RHM_Text, chaikin_lubensky_2000}, heterogeneous materials \cite{RHM_Text}, granular media \cite{mehta_1994}, biological systems \cite{LIANG_bio1,Purohit_bio2,Gevertx_bio3}, and many other physical phenomena; see also refs \citenum{Torquato_JamHardPart} and \citenum{Torquato_PackingPersp}.
A hard-particle packing is a collection of nonoverlapping objects in $d$-dimensional Euclidean space $\mathbb{R}^d$.
The packing fraction $\phi$ is the fraction of space these particles occupy.
A considerable amount of work on such packings in both two \cite{Torquato_JamHardPart, Torquato_PackingPersp, conway_sloane_book, Rogers_Packing, Toth_Somepacking, Atkinson_Noncirc} and three dimensions \cite{Donev_UnusEll, Torquato_ASCNat, Torquato_tet, deGraaf_nonconvex, Torquato_Organize, conway_sloane_book, Torquato_JamHardPart, Torquato_PackingPersp} concerns ordered packings.

Sufficiently slow compression protocols tend to produce the densest packings, which are periodic in low space dimensions \cite{Torquato_JamHardPart, Torquato_PackingPersp, conway_sloane_book}.
Increasing the compression rate can result in defective crystalline, polycrystalline, and even glassy states \cite{Torquato_JamHardPart}, including maximally random jammed (MRJ) states \cite{Torquato_RCPBad}.
Qualitatively, MRJ packings are maximally disordered and mechanically rigid systems that can be used to emulate structural glasses \cite{Torquato_JamHardPart,Torquato_RCPBad}.
Such packings are known to be \textit{hyperuniform}, meaning their infinite-wavelength density fluctuations are anomalously suppressed compared to those in typical disordered systems \cite{Torquato_HUDef, Donev_Unexpected, Torquato_HURev}.

Disorder can occur in dense, jammed monodisperse sphere packings in $\mathbb{R}^3$ as a result of \textit{geometrical frustration}.
A packing is geometrically frustrated if the local densest packing arrangement is incompatible with the global densest packing arrangement.
These sphere packings are geometrically frustrated because the densest local arrangement, which is tetrahedral, cannot optimally fill the space\cite{Jullien_Frust,RHM_Text}, and thus is inconsistent with the global densest arrangements, the FCC lattice and its stacking variants\cite{hales_2005}.
Monodisperse packings of many other particle shapes in $\mathbb{R}^3$ are known to have MRJ states including ellipsoids \cite{Donev_MRJelip}, superballs \cite{Jiao_MRJballs}, the Platonic solids \cite{Jiao_MRJplatonic}, and truncated tetrahedra \cite{Chen_MRJtruntet}.

Monodisperse circular disks in $\mathbb{R}^2$ lack geometrical frustration because their densest local packing is consistent with the densest global packing, which is the triangular lattice \cite{RHM_Text}.
As a result, use of garden-variety compression algorithms produces dense packings that are polycrystalline with a probability of nearly unity\cite{Atkinson_2DMono}.
Bidisperse disks and superdisks in $\mathbb{R}^2$, however, are known to form disordered, jammed packings\cite{Tian_diskMRJ,Speedy_Glass,Donev_GlassTransition}.

Packings of particles in two dimensions are of particular interest because of their ability to model thin films \cite{ohring_2002}, single-layer molecular adsorption onto flat substrates \cite{Azzam_film1,Cyganik_film2}, and two-dimensional biological systems like epithelial cells \cite{FARHADIFAR_ep1,CLASSEN_ep2}. 
Characterization of kinetic effects in such packings can aid in the design of experiments involving packings at interfaces for example, using Langmuir-Blodgett troughs to synthesize monolayers \cite{Dabbousi_QDexpt, BAsavaraj_latexexpt} or to verify the results of experimental two-dimensional packings \cite{Zhao_exptveri}.

It is important to obtain a more complete understanding of kinetic effects due to their ubiquity in numerical and experimental hard-particle packing protocols.
Specifically, we examine the extent to which kinetic effects cause deviations, in packing fraction and order, with respect to the corresponding densest possible configuration of a two-dimensional monodisperse hard-particle packing as a function of the particle shape.

In this work, we use a stochastic search implementation of the adaptive shrinking cell (ASC) optimization scheme \cite{Atkinson_Noncirc,Torquato_ASCNat,Torquato_ASCPRE,Jiao_ASCLP} to generate dense packings of convex, noncircular objects.
In this algorithm, the packing fraction is maximized by sequentially applying random translations and rotations to hard particles in a periodic fundamental cell, which is subsequently isotropically compressed (or dilated) and sheared.
These steps are repeated until the increase in packing fraction is less than a small numerical tolerance, at which point the packing is considered effectively jammed (within some tolerance).
We mimic different effective time scales to examine kinetic effects using this implementation by modulating the number of particle moves between compression/shear steps.
We determine how the degree of rotational symmetry of a particular shape, curvature, and asphericity impact packing kinetics by studying packings of rhombi, obtuse scalene triangles, lenses, curved triangles \cite{Atkinson_Noncirc}, and so-called ``ice cream cones".

To contrast with geometrical frustration, we introduce the idea of \textit{kinetic frustration}, which is the deviation of a packing from its densest possible configuration caused by the compression rate.
We characterize the degree of kinetic frustration using the \textit{kinetic frustration index} $K$ (defined in section \ref{ref:frustdef}), which measures the deviation of a packing from its maximum possible packing fraction. 
We determine the effect of asphericity, curvature, and rotational symmetry on $K$.
To quantify the degree of short- and long-range translational and orientational order in these packings, we compute their \textit{spectral densities} $\spD{\vect{k}}$ (defined in section \ref{sec:SpecDens_def}).
The spectral density is the Fourier transform of the autocovariance of the phase indicator function and can be obtained from scattering experiments \cite{RHM_Text, Debye_Scattering}.
We also determine the type and number of contacts in each packing to within a small numerical tolerance (see section \ref{sec:ASCScheme}) and determine the resulting contact network.
We use the contact networks to determine how the average number of constraints on each particle $Z_C$ (see section \ref{sec:Jam}) and the fraction of particles that can freely move within cages of the jammed backbone, or \textit{rattlers}, vary as a function of compression rate and particle shape.
We expect the examination of such kinetic effects on this large collection of shapes to be relevant to the design of experimental packing protocols.

The rest of the paper is organized as follows. 
Section \ref{sec:background} contains the background pertaining to and the mathematical definitions of the methods used to produce and characterize the dense particle packings.
Section \ref{sec:ASCScheme} describes the ASC scheme and its adaptation used here to generate particle packings with different compression rates.
Section \ref{sec:ShapeDefns} defines the noncircular particle shapes considered in this work as well as information about the densest packings for shapes that do not tile the plane.
In section \ref{sec:Results}, we present results for the kinetic frustration and degree of order/disorder in the final packings.
We then offer conclusions and plans for future research in section \ref{sec:conclusion}.

\section{Computational Details} \label{sec:background}

\subsection{Packings}\label{sec:Packings}

Here, we give the pertinent definitions following closely refs \citenum{Torquato_JamHardPart} and \citenum{Atkinson_Noncirc}. 
A \textit{lattice} $\mathbf{\Lambda}$ in $\mathbb{R}^d$  is a subgroup comprising integer linear combinations of a set of $d$ vectors, $\{\mathbf{p}_1,\mathbf{p}_2,...,\mathbf{p}_d\}$, which are a basis for $\mathbb{R}^d$.
This, in the physical sciences and engineering, is termed a $Bravais$ lattice.
A \textit{packing}, $P$, is a collection of nonoverlapping particles in $\mathbb{R}^d$.
If all members of $P$ are translates of each other, where the vectors of translation form a lattice, $P$ is known as a \textit{Bravais lattice} (or, simply, \textit{lattice}) \textit{packing}.
More generally, if $P$ can be decomposed into a set of \textit{N}$\geq1$ distinct lattice packings all with the same lattice vectors, $P$ is said to be a \textit{periodic packing} with an \textit{N-particle basis}.
A periodic packing with a two-particle basis of particular importance is the \textit{double lattice packing}, examples of which are given in section \ref{sec:DensPackings}.
A packing $P$ is a double lattice packing if $P$ can be decomposed into two lattice packings, $P_0$ and $P_1$, such that an inversion around some point in the space interchanges $P_0$ and $P_1$.
For each of these periodic packings, there is an associated \textit{fundamental cell} $F$, parallelotopic in shape, defined by the lattice matrix $\mathbf{\Lambda}=\{\mathbf{p}_1,\mathbf{p}_2,...,\mathbf{p}_d\}$ containing all \textit{N} particle centroids.
The packing fraction, $\phi$, is the fraction of space that the particles cover.
For a monodisperse periodic packing with an $N$-particle basis $\phi$ is given by
\begin{equation}
    \phi=\frac{Nv_1}{\textrm{Vol}(F)}
\end{equation}
where $v_1$ is the volume of a single $d$-dimensional particle and Vol$(F)$ is the $d$-dimensional volume of the fundamental cell.

\subsection{Kinetic Frustration}\label{ref:frustdef}

The focus of this work is the characterization of dense hard-particle packings generated by using packing protocols with different compression rates. 
Kinetic frustration is the deviation of a packing from its densest possible configuration caused by the compression rate.
To quantify this, we define the kinetic frustration index $K$:
\begin{equation}\label{eq:kfi_def}
    K = 1-\frac{\phi}{\phi_{max}}
\end{equation}
where $\phi_{max}$ is the maximum possible packing fraction associated with a given packing.
Larger values of $K$ indicate larger deviations from the maximum possible packing fraction.
Knowledge of $K$ as a function of particle shape and compression rate may be relevant to the design of laboratory packing protocols.
Additionally, we compute the spectral density (defined below) of each packing to quantify the degree of short- and long-range ordering and determine how ordering is correlated to $K$.

\subsection{Jamming Categories and Isostaticity}\label{sec:Jam}
\textit{Jammed} hard-particle packings are those in which mechanical stability of a specific type is conferred to the packing via interparticle contacts \cite{Torquato_JamHardPart}.
Three broad and mathematically precise jamming categories can be distinguished based on the nature of the mechanical stability conferred, which in order of increasing stability are as follows \cite{Torquato_HPPDefs,Torquato_JamHardPart}:
(1) \textit{Local jamming}: no individual particle can be moved while holding all other particles fixed.
(2) \textit{Collective jamming}: the packing is locally jammed, and no collective motion of a finite subset of particles is possible. 
(3) \textit{Strict jamming}: the packing is collectively jammed and all volume-nonincreasing deformations are disallowed by the impenetrability constraint.
A special jammed state is the \textit{maximally random jammed} (MRJ) state, which is defined as the most disordered configuration (as measured by a set of scalar order metrics) subject to a particular jamming category \cite{Torquato_RCPBad}.
Such packings are known to possess hyperuniform density fluctuations \cite{Donev_Unexpected, Zachary_QLRLet,Zachary_QLRI,Zachary_QLRII}.

Jammed packings are also characterized by the number of constraints imposed by interparticle contacts relative to the number of degrees of freedom (DOF) in the packing.
In hard disk or ellipse packings, for example, each interparticle contact occurs at a single point and corresponds to a single constraint.
Faceted particles can have contacts at more than a single point, e.g., edge-to-edge contacts or face-to-face contacts, which impose additional constraints by blocking rotations and must be weighted accordingly when counting the number of constraints \cite{Jaoshvili_TetDice}.
MRJ sphere packings in $\mathbb{R}^3$, for example, are known to be \textit{isostatic} \cite{ALEXANDER_Amorphous, Edwards_StatMechStress, Donev_NearJam, Ohern_Epitome}, meaning that total number of constraints is equal to the number of DOF in the system.
The number of constraints required for isostaticity will depend on the particle shape, jamming category, and boundary conditions.
For example, an isostatic strictly jammed disk packing in $\mathbb{R}^2$ under periodic boundary conditions must have $2N+1$ constraints, where $N$ is the number of particles \cite{Donev_NearJam}.
In the infinite-particle-number limit, the average number of constraints per particle $Z_C$ in an isostatic packing is equal to twice the number of degrees of freedom per particle $f$ (i.e., $Z_C=2f$) \cite{Jiao_MRJplatonic, Torquato_JamHardPart}.
Packings with more constraints than isostatic ones are \textit{hyperstatic}, and those having fewer constraints are \textit{hypostatic}.
While MRJ sphere packings in three dimensions \cite{ALEXANDER_Amorphous, Edwards_StatMechStress, Donev_NearJam, Ohern_Epitome} and MRJ disk packings in two dimensions \cite{Atkinson_2DMono} are isostatic, this is not generally true of all MRJ packings.
In particular, certain aspherical particles with smooth boundaries, like ellipsoids \cite{Donev_MRJelip}, superellipsoids \cite{Delaney_SueprMRJ}, and superballs \cite{Jiao_MRJballs}, have hypostatic MRJ packings. 
In the two-dimensional packings considered herein, edge-to-edge contacts impose two constraints on each particle, and all other contact types (e.g., vertex-to-edge) impose one constraint on each particle.

In practice, jammed hard-particle packings produced via simulations or experiments contain a small concentration of \textit{rattlers}, which are not jammed but are locally imprisoned by neighboring jammed particles \cite{Torquato_JamHardPart, Torquato_RCPBad}.
None of the three jamming definitions above permit the presence of rattlers.
Nevertheless, it is the significant majority of hard particles that confers rigidity to the packing, and in any case, the rattlers could be removed (in computer simulations) without disrupting the remaining jammed particles \cite{Torquato_JamHardPart}.
The \textit{rattler fraction}, $\phi_R$, is greatest in sphere packings in any dimension and is significantly decreased when rotational degrees of freedom are introduced \cite{Jiao_MRJballs} or the spatial dimension is decreased \cite{Atkinson_2DMono}.
Atkinson et al. \cite{Atkinson_Slowdown} have shown that removal of rattlers from MRJ packings results in a nonhyperuniform packing, meaning that the subset of jammed particles alone is far from hyperuniform.
In this work we consider packings that do not have their rattlers removed.

\subsection{Spectral Density}\label{sec:SpecDens_def}
A hard-particle packing can be modeled as a two-phase heterogeneous medium, where the matrix phase $\mathcal{V}_1$ is the void space between the particles and the particle phase $\mathcal{V}_2$ is the space occupied by the particles, such that $\mathcal{V}_1\cup\mathcal{V}_2=V\subset\mathbb{R}^d$. \cite{Cinacchi_Lenses}.
The (micro)structure of the packing can be fully characterized by a countably infinite set of \textit{$n$-point probability functions} $S_n^{(i)}$, defined by \cite{RHM_Text}
\begin{equation}
    S_n^{(i)}(\mathbf{x}_1,\dots,\mathbf{x}_n)=\left\langle\prod^n_{j=1}\mathcal{I}^{(i)}(\mathbf{x}_n)\right\rangle,
\end{equation}
where $\mathcal{I}^{(i)}$ is the indicator function for phase $i$:
\begin{equation}
    \mathcal{I}^{(i)}(\mathbf{x}) =
    \begin{cases}
    1, & \mathbf{x} \in \mathcal{V}_i\\
    0, & \textrm{else}.
    \end{cases}
\end{equation}
The function $S_n^{(i)}$ gives the probability of finding $n$ points at positions $\mathbf{x}_1,\dots,\mathbf{x}_n$ in phase $i$.
In what follows, we drop the superscript $i$, and restrict our discussion to $\mathcal{V}_2$.

For statistically homogeneous media, $S_n(\mathbf{x}_1,\dots,\mathbf{x}_n)$ is translationally invariant and, in particular, the one-point correlation function is independent of position and equal to the packing fraction
\begin{equation}
    S_1(\mathbf{x})=\phi,
\end{equation}
while the two-point correlation function $S_2(\mathbf{r})$ depends on the displacement vector $\mathbf{r}\equiv\mathbf{x}_2-\mathbf{x}_1$.
The corresponding two-point autocovariance function $\chi_V(\mathbf{r})$ \cite{RHM_Text,StochaticGeo_Text, Quint_Autoco} is obtained by subtracting the long-range behavior from $S_2(\mathbf{r})$:
\begin{equation}
    \chi_V(\mathbf{r})=S_2(\mathbf{r})-\phi^2 
\end{equation}
The nonnegative \textit{spectral density}, $\spD{\vect{k}}$, is defined as the Fourier transform of $\chi_V(\mathbf{r})$ \cite{RHM_Text}, i.e.,
\begin{equation}
    \spD{\vect{k}} = \int_{\mathbb{R}^d}\chi_V(\mathbf{r})e^{-i\mathbf{k}\cdot\mathbf{r}}d\mathbf{k}.
\end{equation}
A hyperuniform packing is one in which  $\spD{\vect{k}} \rightarrow 0$ as $\vect{k}\rightarrow0$ \cite{Zachary_2phaseHU}.

Because of the rigidity of MRJ packings and the presence of a well-defined contact network, Torquato and Stillinger conjectured that any strictly jammed and \textit{saturated} packing is hyperuniform \cite{Torquato_HUDef}.
A \textit{saturated} packing is one in which there is no space available to add another particle of the same kind to the packing.
Subsequently, Zachary et al. \cite{Zachary_QLRI,Zachary_QLRII,Zachary_QLRLet} have shown that MRJ packings of hard particles with shape and size distributions possess vanishing infinite-wavelength local-volume-fraction fluctuations and signature \textit{quasi-long-range} (QLR) pair correlations.
These QLR correlations are manifested by a \textit{linear} scaling in the small-wavenumber region of  $\spD{\vect{k}}$, i.e. $\spD{\vect{k}} \sim \mathbf{k}$ as $\mathbf{k} \rightarrow 0$.

In the present work, we consider packings of $N$ hard particles within a fundamental cell under periodic boundary conditions.
Under these conditions, we can express the spectral density $\spD{\vect{k}}$ of a finite hard-particle packing as \cite{Zachary_QLRI}
\begin{equation}
\begin{split}
    \spD{\vect{k}} = &\frac{\left|\sum_{j=1}^N\textrm{exp}(-i\mathbf{k}\cdot\mathbf{r}_j)\tilde{m}(\mathbf{k};\mathbf{R}_j)\right|^2}{V}\\
    &(\mathbf{k}\neq0),
\end{split}
\end{equation}
where $\{\mathbf{r}_j\}$ denotes the set of particle centroids, $\mathbf{R_j}$ denotes all of the geometrical parameters of the particle shape, $V$ is the volume of the simulation box (fundamental cell), and $\tilde{m}(\mathbf{k};\mathbf{R_i})$ is the Fourier transform of the particle indicator function defined as
\begin{equation}
    m(\mathbf{r};\mathbf{R_i}) =
    \begin{cases}
    1, & \mathbf{r} \textrm{ is in particle }i\\
    0, & \textrm{otherwise.}
    \end{cases}
\end{equation}
The shape of the fundamental cell, defined by the lattice vectors \{$\mathbf{p}_i\}$, restricts the wavevectors such that $\mathbf{k}\cdot\mathbf{p}_i=2\pi n\;\forall\;i$, where $n\in\mathbb{Z}$.
Visualization of the spectral densities allows us to examine the degree of short- and long-range translational and orientational order in the packings.
Instead of computing $\tilde{m}(\mathbf{k};\mathbf{R_i})$ for each particle shape, we take the Fourier transform of a square pixelization of the packing, requiring only $\tilde{m}(\mathbf{k};\mathbf{R_i})$ for a square (see, e.g., Ref. \citenum{CHEN_pixel}).
The angular-averaged spectral densities in this work are ensemble averaged over 50 configurations.
Accompanying two-dimensional spectral densities are for a single, representative configuration in order to account for important orientational information in anisotropic packings, which may be lost upon angular averaging.

\subsection{Adaptive Shrinking Cell (ASC) Optimization \\Scheme}\label{sec:ASCScheme}
The Torquato-Jiao ASC scheme generates dense hard-particle packings in a periodic fundamental cell by translating and rotating the particles while simultaneously shrinking and deforming the boundary of the fundamental cell.
This process can be formally stated as follows\cite{Atkinson_Noncirc}:
\begin{equation}
\begin{split}
    &\textrm{minimize}:\;-\phi(\mathbf{r}_1^{\mathbf{p}},\mathbf{r}_2^{\mathbf{p}},\mathbf{r}_3^{\mathbf{p}},\dots,\mathbf{r}_N^{\mathbf{p}};\\
    &\;\;\;\;\;\;\;\;\;\;\;\;\;\;\;\;\;\;\;\;\;\;\;\;\theta_1,\theta_2,\theta_3,\dots,\theta_N;\Lambda),\\
    &\textrm{such that}: (S_i\cap S_j) \subseteq (\Gamma_i\cup\Gamma_j)\\
    &\;\;\;\;\;\;\;\;\;\;\;\;\;\;\;\;\;\;\;\;\forall\; i,j=1,2,3,\dots,N,\;i\neq j
\end{split}
\end{equation}
where $N$ is the number of particles, $\mathbf{r}_i^{\mathbf{p}}$ and $\theta_i$ denote the position and orientation of particle $i$, respectively, $S_i$ is the closed set in $\mathbb{R}^2$ associated with particle $i$, and $\Gamma_i$ is the boundary of the set $S_i$.
This optimization scheme can be solved by using either stochastic \cite{Torquato_ASCNat,Torquato_ASCPRE,Atkinson_Noncirc} or linear programming \cite{Jiao_ASCLP} techniques. 
This work uses the Monte Carlo implementation described in ref \citenum{Atkinson_Noncirc}.

\begin{figure}[t]
        \subfloat[]{{\label{fig:ASC_1}}
            \includegraphics[width=0.15\textwidth]{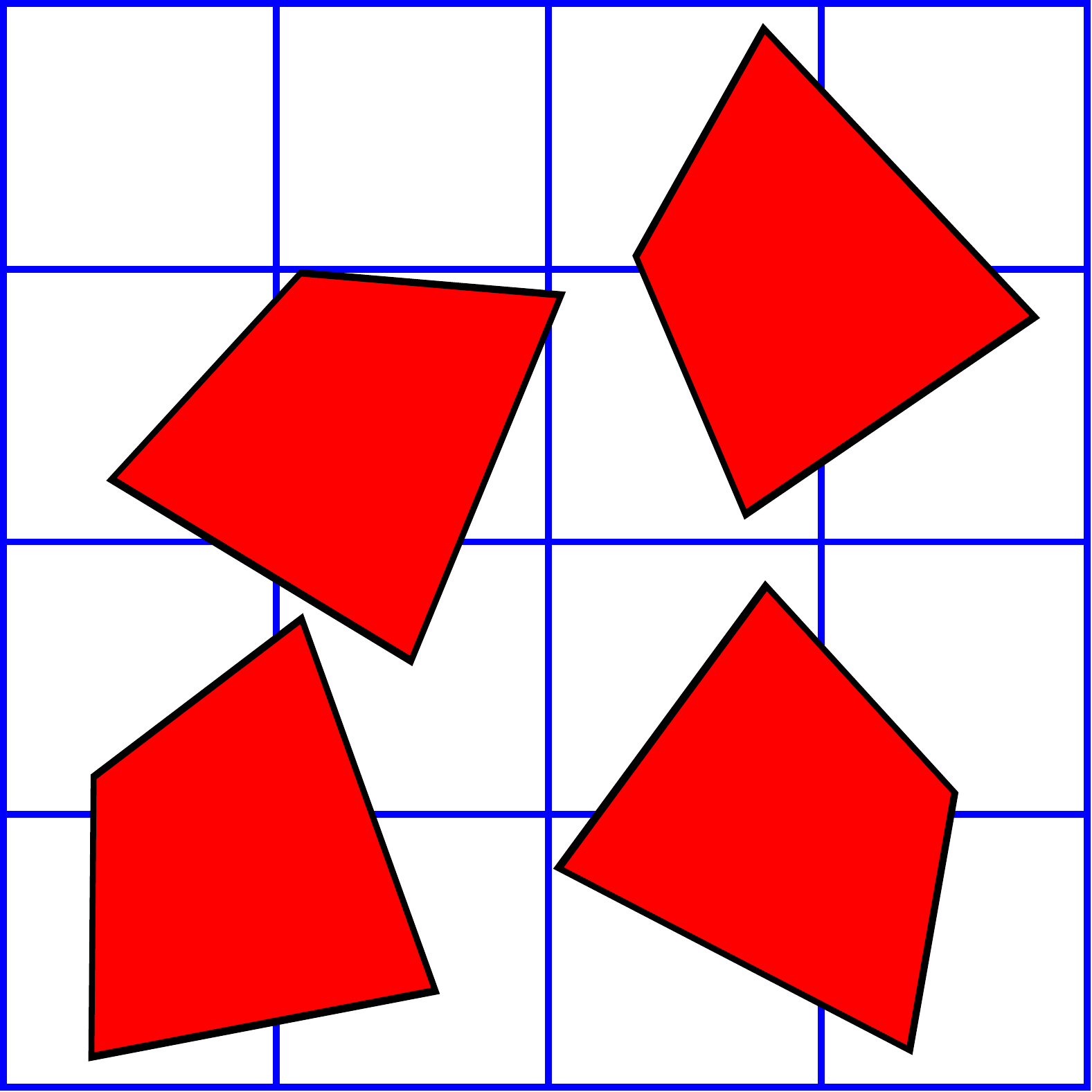}}
        \subfloat[]{{\label{fig:ASC_2}}
            \includegraphics[width=0.15\textwidth]{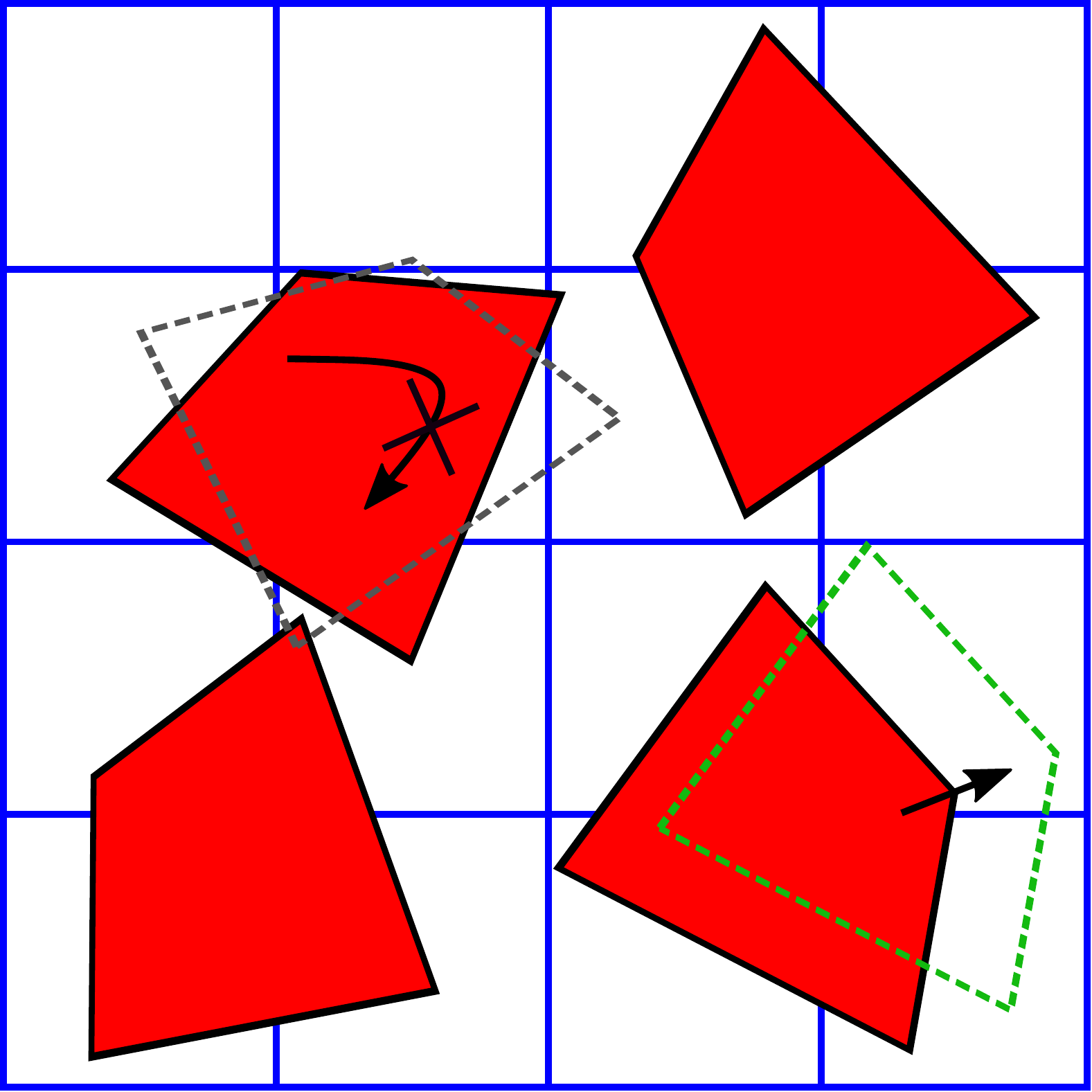}}
        \subfloat[]{{\label{fig:ASC_3}}
            \includegraphics[width=0.15\textwidth]{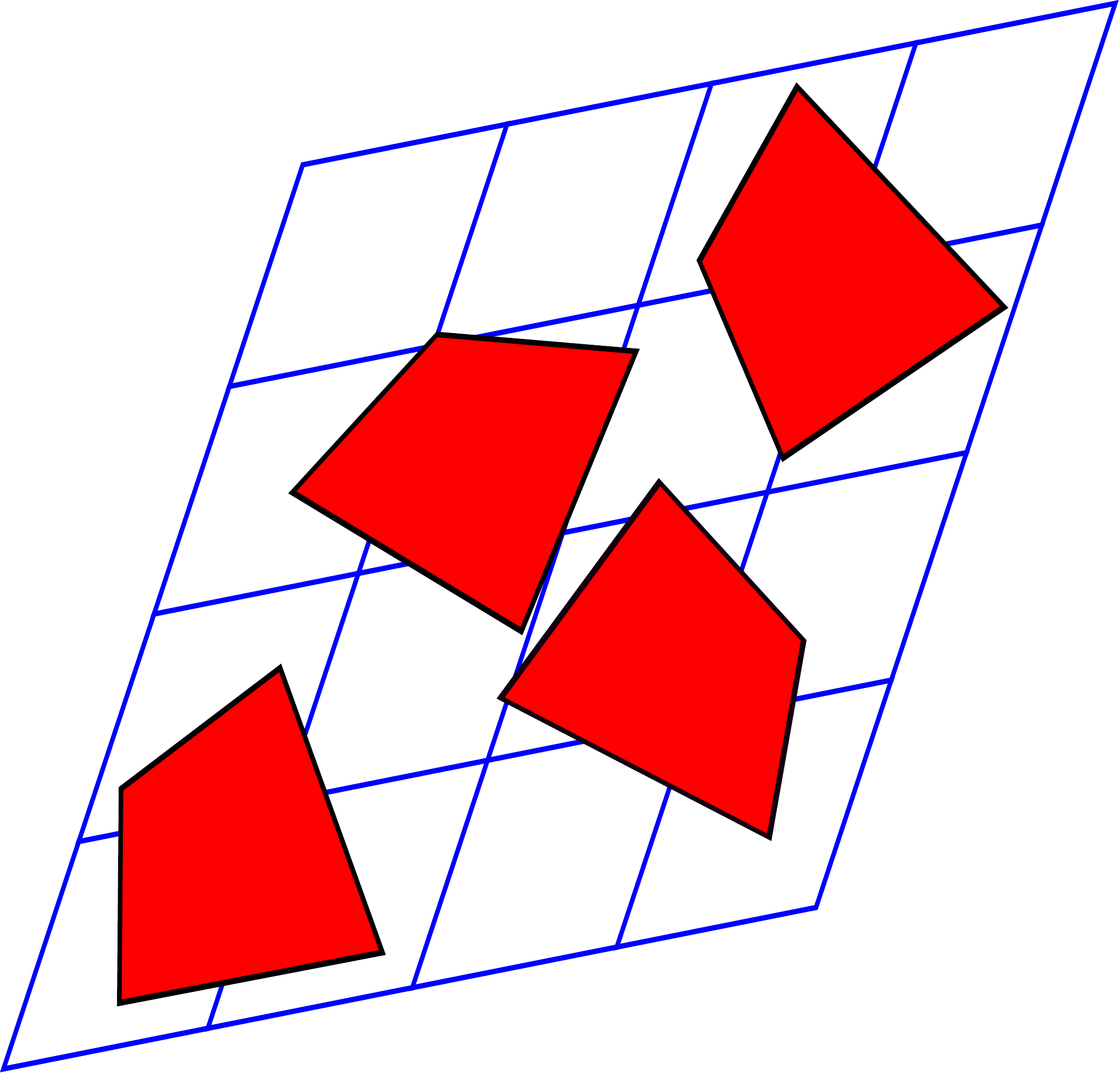}}    
         \caption{Schematic of a single step of the ASC optimization scheme. (a) The initial configuration of the 4 particles. (b)Two trial moves: a rejected rotation due to violation of the nonoverlapping constraints, and an accepted translation. (c) The result of an accepted trial deformation and compression of the packing.}
        \label{fig:images}
    \end{figure}

The stochastic implementation of this scheme in $\mathbb{R}^2$ uses a periodic parallelogrammatic fundamental cell. 
All $N$ particle centroids are given in \textit{lattice coordinates}, which are relative coordinates with respect to the lattice vectors, i.e., $\mathbf{r}_i^{\mathbf{p}}\in [0, 1)^2$.
The particle orientations are specified by a rotation of a given angle, i.e., $\theta_i\in[0, 2\pi)$.
Given an initial configuration, the stochastic search method uses an iterative process to increase the packing fraction by using the following steps, shown schematically in Figure \ref{fig:images}:
(1) ``Random moves" - Random rotations or translations are applied to the particles.
These moves are only accepted if the resulting configuration satisfies the nonoverlap constraints.
(2) ``Random strains" - A random strain comprising a deformation and dilation or compression is applied to the simulation box.
This strain will either increase or decrease the area of the fundamental cell with some specified probability, corresponding to uphill or downhill moves, respectively.

During the ``random moves" step, a trial translation or rotation with a prescribed maximum magnitude is applied to each of the $N$ particles a set number of times. 
A trial move is accepted if the new particle position does not overlap any of the other particles or their periodic images.
Otherwise, the particle is returned to its previous position or orientation.
The maximum magnitude of these trial movements is steadily decreased throughout the execution of the algorithm by reducing this maximum by some constant ratio when the number of moves accepted per ``random move" step falls significantly below 50\%.

During the ``random strains" step, the simulation box is both deformed and compressed or dilated simultaneously such that the area of the box decreases on average, while ensuring the nonoverlap constraint remains satisfied.
These strains effect collective motions on the particles because the centroid positions of the particles are defined in terms of the lattice vectors. 
The maximum magnitude of a trial strain and the maximum number of trials are prescribed.
The first trial that does not violate the nonoverlap constraints is accepted. 
After each unsuccessful trial, the maximum magnitude of the strain is decreased until either a trial is accepted or the maximum number of trials has been attempted.
When either of these conditions is met, we return to the first step of the scheme and the maximum magnitude of the strain is reset to the value from the beginning of the step. 

The linear programming solution to the ASC scheme was designed to produce the \textit{inherent structure} associated with an initial, unjammed, sphere packing \cite{Jiao_ASCLP}.
An inherent structure is a minimum in the energy landscape \cite{Stillinger_Hiddenstruct,Stillinger_InherentStruct}.
For packings, this corresponds to a mechanically rigid and locally maximally dense configuration \cite{Stillinger_InherentPacking,Jiao_ASCLP}.
These inherent structures vary in their packing fraction and degree of order, and the number of possible inherent structures increases with $N$.
The linear programming solution given by Torquato and Jiao in ref \citenum{Jiao_ASCLP} guarantees with a high probability the generation of a jammed sphere packing across dimensions for $d\geq2$, with a wide variety of packing fractions and degrees of order.
An inherent structure is highly dependent on its initial configuration, and thus unusual initial configurations in principle allow one to obtain unusual inherent structures.

In the present work we apply the ASC scheme to monodisperse packings of 504 particles.
To minimize the effect of the initial conditions, we use a random sequential addition process \cite{RHM_Text} to place particles in a square box with random orientations such that the initial $\phi$ is several orders of magnitude smaller than $\phi_{max}$.
Simulations are terminated when $\phi$ increases less than $10^{-10}$ over the course of 100 ``random strain" steps, at which point it is assumed $\phi$ has reached a local maximum.
To impose different effective time scales (compression rates), we use ``slow", ``medium", and ``fast" compression schedules, in which there are 1000, 100, and 10 trial moves per particle in the ``random moves" step, respectively. 
The three schedules have identical movement and strain magnitudes and move strictly downhill (i.e., no ``random strain" step increases the area of the simulation box).
To generate the contact network, we take the dense output from the procedure above and use a fourth ``contact generation" schedule with much smaller movement and compression/shear magnitudes.
This fourth schedule is terminated when the interparticle distances (excluding rattlers) are smaller than $10^{-10}D_0$ for particle shapes composed of arcs and $6\times10^{-10}D_0$ for particle shapes containing flat edges, where $D_0$ is the largest distance from the particle centroid to its boundary, at which point the packing is considered effectively jammed.  
We use a coarser tolerance for faceted particles due to the increased computational cost associated with the additional rotational constraints imposed by flat edges as jamming is approached.
Kinetic frustration and contact network results for each combination of particle shape (see section \ref{sec:ShapeDefns}) and compression schedule are averaged over three configurations.

\subsection{Particle Shapes}\label{sec:ShapeDefns}

Here, we mathematically define the particle shapes studied in this work.
Additionally, we state the maximum packing fractions $\phi_{max}$ of these shapes, which are required to compute the kinetic frustration index $K$ (cf. eq \ref{eq:kfi_def}), as well as relevant geometrical properties.
One such property is the asphericity $\gamma$ defined as \cite{Torquato_ASCNat,Torquato_ASCPRE}
\begin{equation}
    \gamma = \frac{r_{\textrm{smallest bounding circle}}}{r_{\textrm{largest inscribable circle}}}
\end{equation}
We examine obtuse scalene triangles, rhombi, and curved triangles.
Each of these shapes has a known $\phi_{max}$.
We also study lenses and ``ice cream cones." 
To our knowledge the $\phi_{max}$ of these shapes is unknown.

\subsubsection{Shape Definitions}\label{sec:ShapeMathDefs}
\begin{figure*}[th]
        \centering
        \subfloat[]{\label{fig:Rhomb_temp}
        \includegraphics[width = 0.3\textwidth]{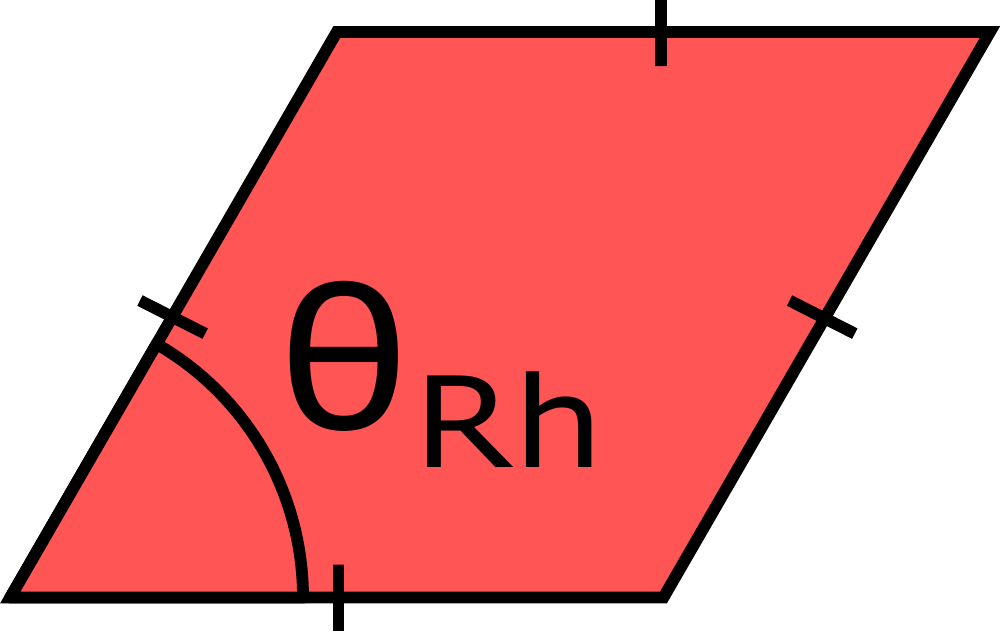}
        }
        \subfloat[]{\label{fig:OST_temp}
        \includegraphics[width = 0.3\textwidth]{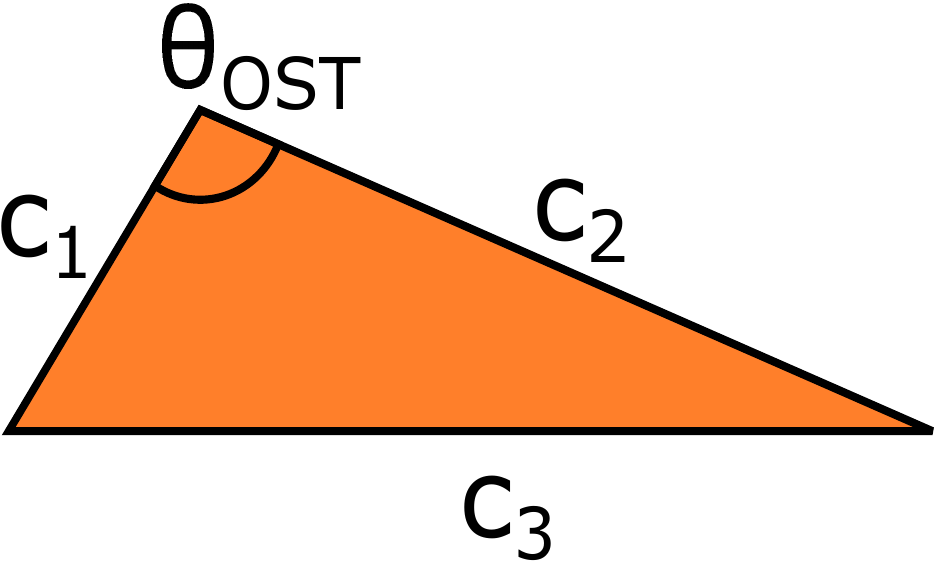}
        }
        \subfloat[]{\label{fig:CT_temp}
        \includegraphics[width = 0.3\textwidth]{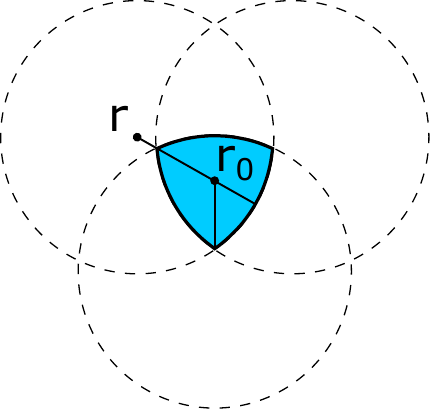}
        }
        \vspace{0.05\textheight}
        \smallskip
        \subfloat[]{\label{fig:Lens_temp}
        \includegraphics[width = 0.3\textwidth]{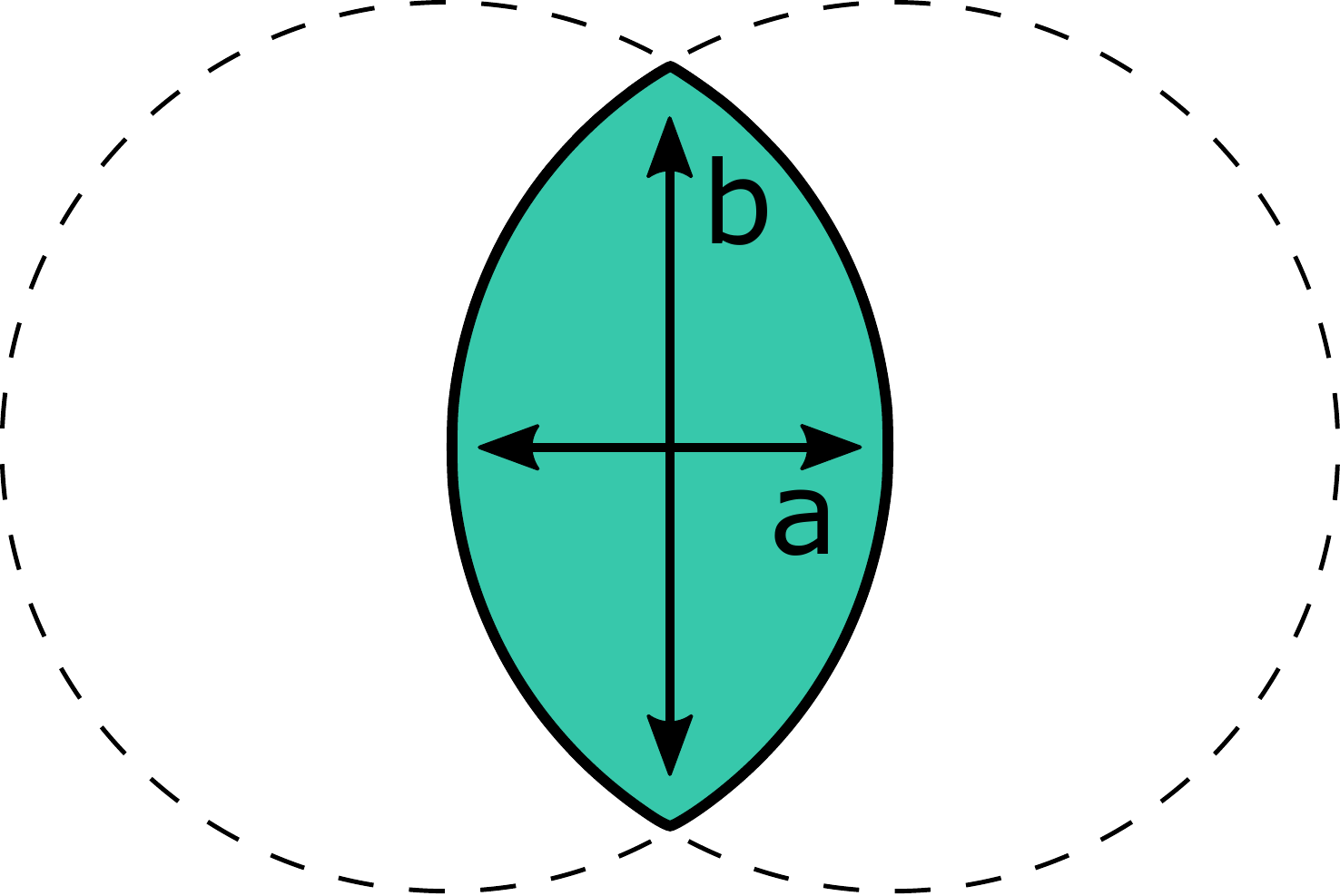}
        }
        \subfloat[]{\label{fig:ICC_temp}
        \includegraphics[width = 0.3\textwidth]{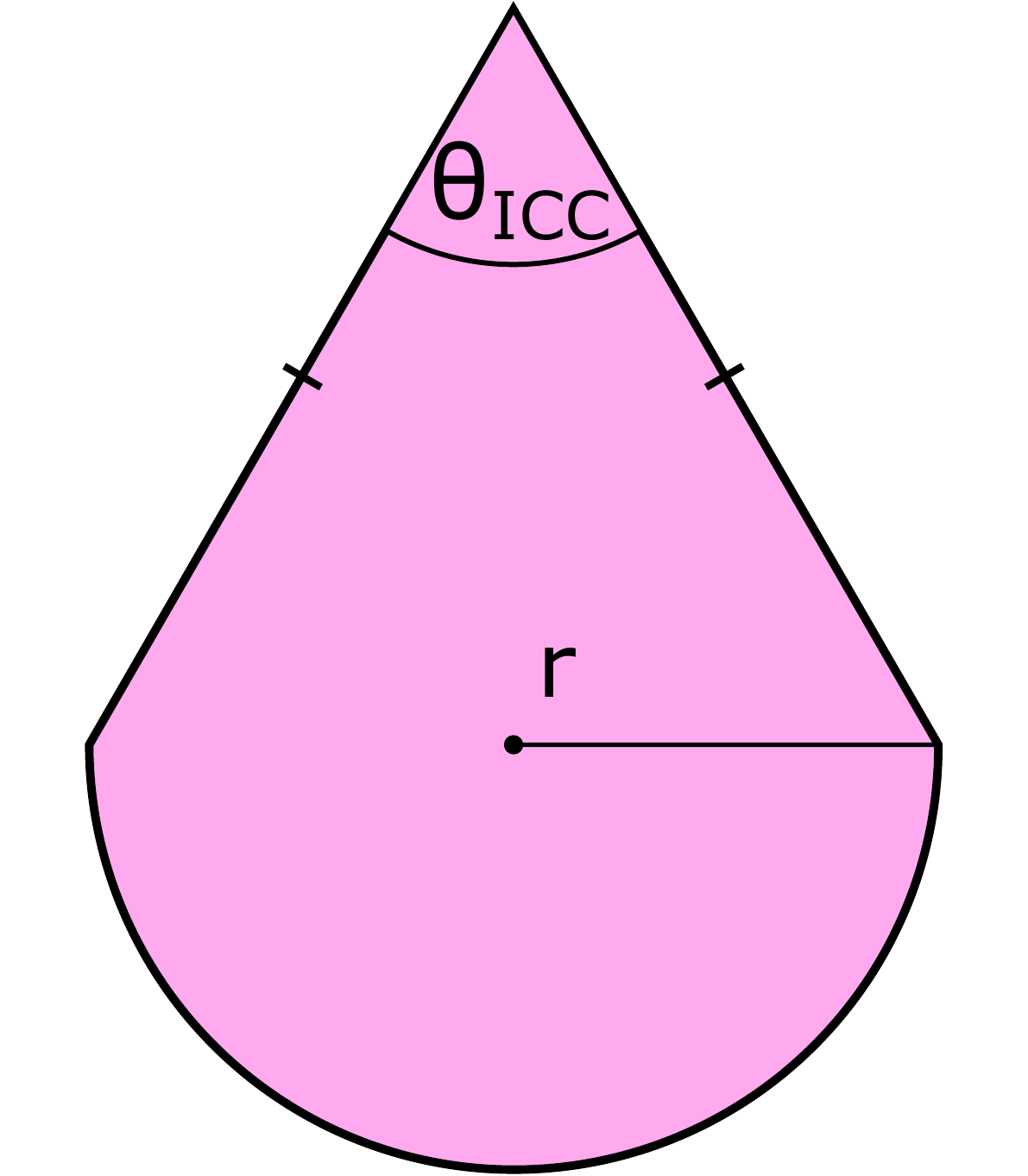}
        }
        \subfloat[]{\label{fig:Bow_temp}
        \includegraphics[width = 0.3\textwidth]{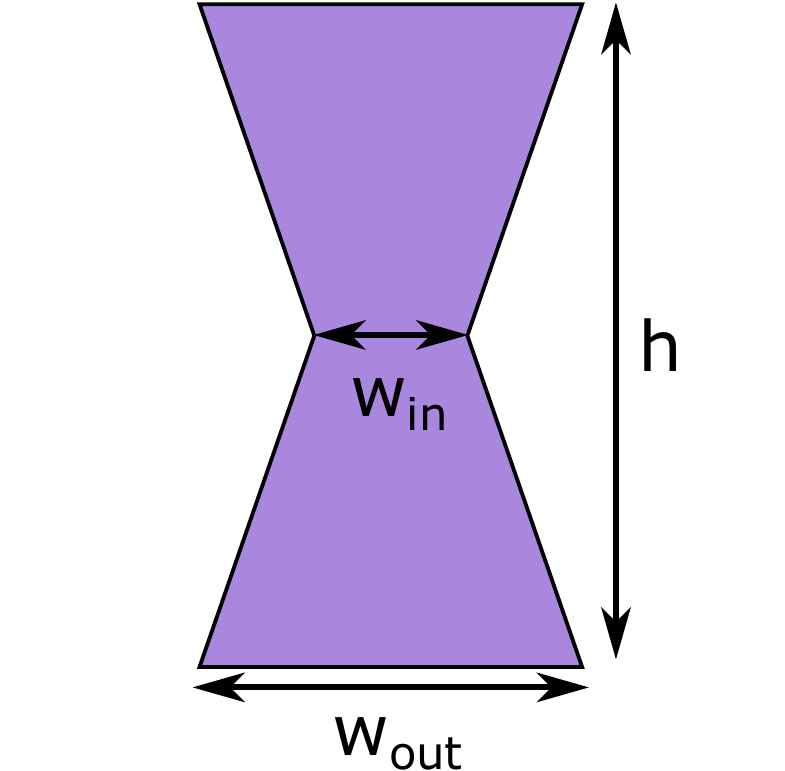}
        }
        \caption{Geometrical properties of the (a) rhombi, (b) obtuse scalene triangles, (c) curved triangles, (d) lenses, (e) ice cream cones, and (f) bowties examined in this work. Shapes (a) - (e) are the convex bodies examined in the main body of this work, and object (f) is a concave body discussed briefly in section \ref{sec:conclusion}. Objects (a), (b), and (f) tile the plane, while (c) - (e) do not. Only (b) is chiral, and racemic mixtures of this object are discussed briefly in section \ref{sec:conclusion}.}
        \label{fig:Shape_prot}
        \end{figure*} 

\paragraph{Rhombi.}\label{sec:Rhombs}
A rhombus is a quadrilateral with four equal sides, an example of which is given in Figure \ref{fig:Shape_prot}a.
The angle $\theta_{Rh} \in (0^{\circ},90^{\circ}]$, generating rhombi that interpolate between a thin line and a square.
Rhombi have $D_2$ symmetry, except for in the square limit, where they have $D_4$ symmetry.
We generate packings of rhombi with $\theta_{Rh}=\{30,40,50,60,70,80\}$, which have $\gamma = \{3.86, 2.92, 2.37, 2.00, 1.74, 1.56\}$, respectively.
$\phi_{max}=1$ for all rhombi.

\paragraph{Obtuse Scalene Triangles.}\label{sec:OST}
An obtuse scalene triangle (OST) has three unequal sides and one angle greater than 90$^{\circ}$, an example of which is given in Figure \ref{fig:Shape_prot}b.
We base our OSTs on the \textit{first candidate} from ref \citenum{Robin_Scalene}, which has angles of 18.6$^{\circ}$, 45$^{\circ}$, and 116.4$^{\circ}$. 
Here, we chose that the ratio of the lengths of sides $C_1$ and $C_2$ is 2.216, as they are in the obtuse \textit{first candidate} and vary $\theta_{OST}\in(90^{\circ},180^{\circ})$.
All OST have $C_1$ symmetry.
All simulations containing OSTs are enantiopure, except for those in section \ref{sec:conclusion}.
We generate packings of OSTs with $\theta_{OST}=\{100,110,120,130,140\}$ which have $\gamma = \{3.44, 3.89, 4.51, 5.37, 6.73\}$, respectively.
$\phi_{max}=1$ for all OSTs.

\paragraph{Curved Triangles.}\label{sec:CT}
A curved triangle (CT) is the convex intersection of three congruent circles placed at the vertices of an equilateral triangle \cite{Atkinson_Noncirc}, an example of which is given in Figure \ref{fig:Shape_prot}c.
CTs are characterized by a curvature parameter, $\kappa$, given by
\begin{equation}
    \kappa=\frac{r_0}{r}
\end{equation}
where $r_0$ is the circumradius of the CT and $r$ is the radius of the overlapping circles.
Thus, CTs interpolate between a circle at $\kappa=1$ and an equilateral triangle at $\kappa=0$.
At $\kappa=1/\sqrt{3}$ the particle shape is the same as the well-known Reuleaux triangle \cite{reuleaux_kinematics}.
CTs have $D_3$ symmetry.
We generate packings of CTs with $\kappa=\{1/5, 7/20, 1/2, 1/\sqrt{3}, 13/20, 4/5\}$ which have $\gamma = \{1.74, 1.58, 1.43, 1.37, 1.30, 1.18\}$, respectively.
CTs do not tile the plane, and their $\phi_{max}$ is given in in ref \citenum{Atkinson_Noncirc}.

\paragraph{Lenses.}\label{Sec:Lens}
A lens is the intersection of two congruent circles, an example of which is given in Figure \ref{fig:Shape_prot}d.
Lenses are characterized by an aspect ratio, $\alpha$, given by
\begin{equation}
    \alpha = \frac{a}{b}
\end{equation}
where $a$ is the minor axis and $b$ is the major axis.
Lenses have $D_2$ symmetry, except in the case of the $\alpha = 1$ limit, where the circle has O(2) symmetry.
We generate packings of lenses with $\alpha = \{4/5,2/3,1/2,1/3,1/5\}$ which have $\gamma = \{1.25, 1.5, 2,3,5\}$, respectively.
The $\phi_{max}$ of lenses is, to our knowledge, undetermined.
We give the putative $\phi_{max}$ and corresponding structure in section \ref{sec:DensPackings}.

\paragraph{Ice Cream Cones.}\label{Sec:ICC}
An ``ice cream cone" (ICC) is an isosceles triangle with a semicircle grafted onto its base, an example of which is given in Figure \ref{fig:Shape_prot}e.
The diameter of this semicircle is equal to the length of the base of this isosceles triangle, and the union of the two regions is convex.
The angle $\theta_{ICC}\in(0^{\circ},180^{\circ}]$, generating ICCs that interpolate between an infinitely tall isosceles triangle in the $\theta_{ICC}\rightarrow0^{\circ}$ limit and a semicircle when $\theta_{ICC}=180^{\circ}$.
ICCs have $D_1$ symmetry.
We generate packings of ICCs with $\theta_{ICC}=\{30^{\circ},60^{\circ},90^{\circ},120^{\circ},150^{\circ},180^{\circ}\}$ with $\gamma=\{2.43,1.5,1.21,1.37,1.60,2\}$, respectively.
The $\phi_{max}$ of ICC is, to our knowledge, undetermined.
We give the putative $\phi_{max}$ and corresponding structures in section \ref{sec:DensPackings}.

\begin{figure}[t]
            \centering
            \includegraphics[width=0.4\textwidth]{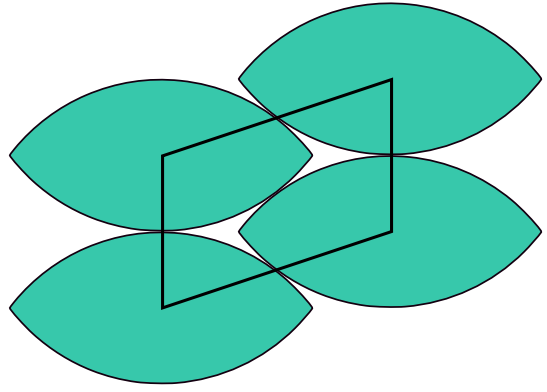}
            \caption{Fundamental basis of the putative densest $\alpha=0.5$ lens packing.}
            \label{fig:LensDensePack}
\end{figure}

\paragraph{Bowties.}\label{Sec:Bowtie}
``Bowties" are rectangles with congruent isosceles triangles taken out of both long sides such that the resulting shape is an irregular hexagon; this is illustrated in Figure \ref{fig:Shape_prot}f.
Bowties are not discussed in the main body of this work due to our focus on convex shapes but have their kinetics discussed briefly in section \ref{sec:conclusion}.
We characterize bowties with a thickness parameter $\beta$, given by
\begin{equation}
    \beta=\frac{w_{in}}{w_{out}}.
\end{equation}
In principle, the aspect ratio of the bowtie can be arbitrary, but here we choose it such that $h = \sqrt{3}w_{out}$. 
Thus, the bowties interpolate between a rectangle when $\beta=1$ and two equilateral triangles attached at a vertex when $\beta=0$.
All bowties herein have $D_2$ symmetry, and $\phi_{max}=1$.

\begin{figure}[t]
                \centering
                \includegraphics[width=0.4\textwidth]{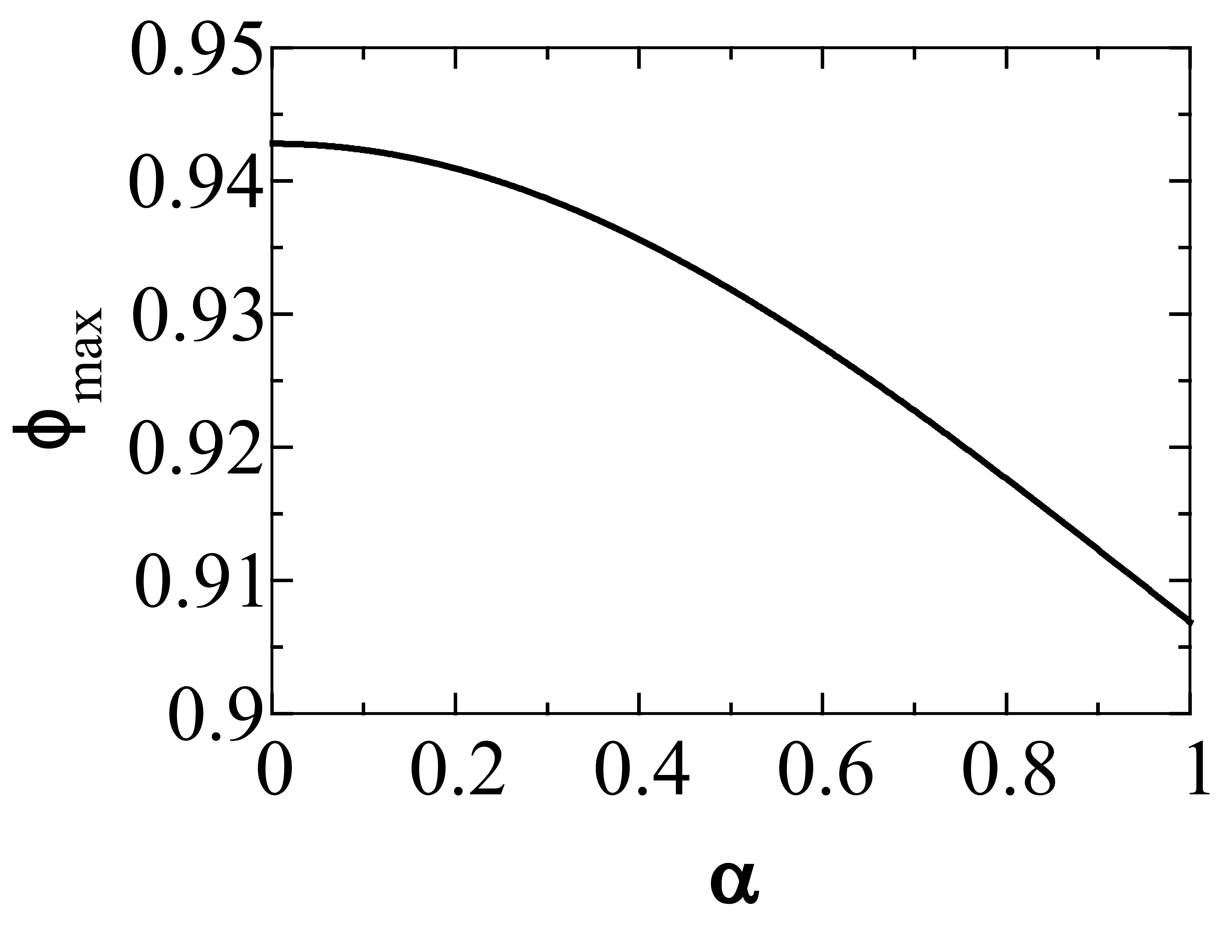}
                \caption{Putative maximum packing fraction $\phi_{max}$ as a function of $\alpha$ for the dense packings of lenses.}
                \label{fig:Lens_DenseFunction}
\end{figure}

\subsubsection{Putative Densest Packings}\label{sec:DensPackings}
    
Here, we follow the procedure given in ref \citenum{Atkinson_Noncirc} to generate dense periodic packings of lenses and ICC to inform analytical predictions for the $\phi_{max}$ of these shapes and the corresponding structures.
We present the smallest periodic repeat units, or \textit{fundamental bases}, of these dense packings, as well as the analytically determined $\phi_{max}$ as a function of the relevant geometrical parameter ($\alpha$ for lenses, $\theta_{ICC}$ for ICC).

\paragraph{Lenses}\label{Sec:DenseLens}    

The putative densest arrangement of lenses is consistent with Fejes T\'{o}th's theorem that the densest packing of a centrally symmetric convex particle is a lattice packing \cite{Toth_Somepacking}.
Figure \ref{fig:LensDensePack} shows the fundamental basis for lenses with $\alpha = 0.5$.
The analytically derived $\phi_{max}$ for lenses is given in Figure \ref{fig:Lens_DenseFunction}.
The $\phi_{max}$ has a maximum of $\frac{2\sqrt{2}}{3}$ in the $\alpha\rightarrow0$ limit and decreases monotonically as a function of $\alpha$.
We find that this packing is consistent with the triangular lattice in the limit of $\alpha = 1$ (circles) with $\phi_{max} = \frac{\pi\sqrt{3}}{6}$.

\begin{figure}[t]
    \centering
            \subfloat[]{\label{fig:T1_ex}
                \includegraphics[width=0.225\textwidth]{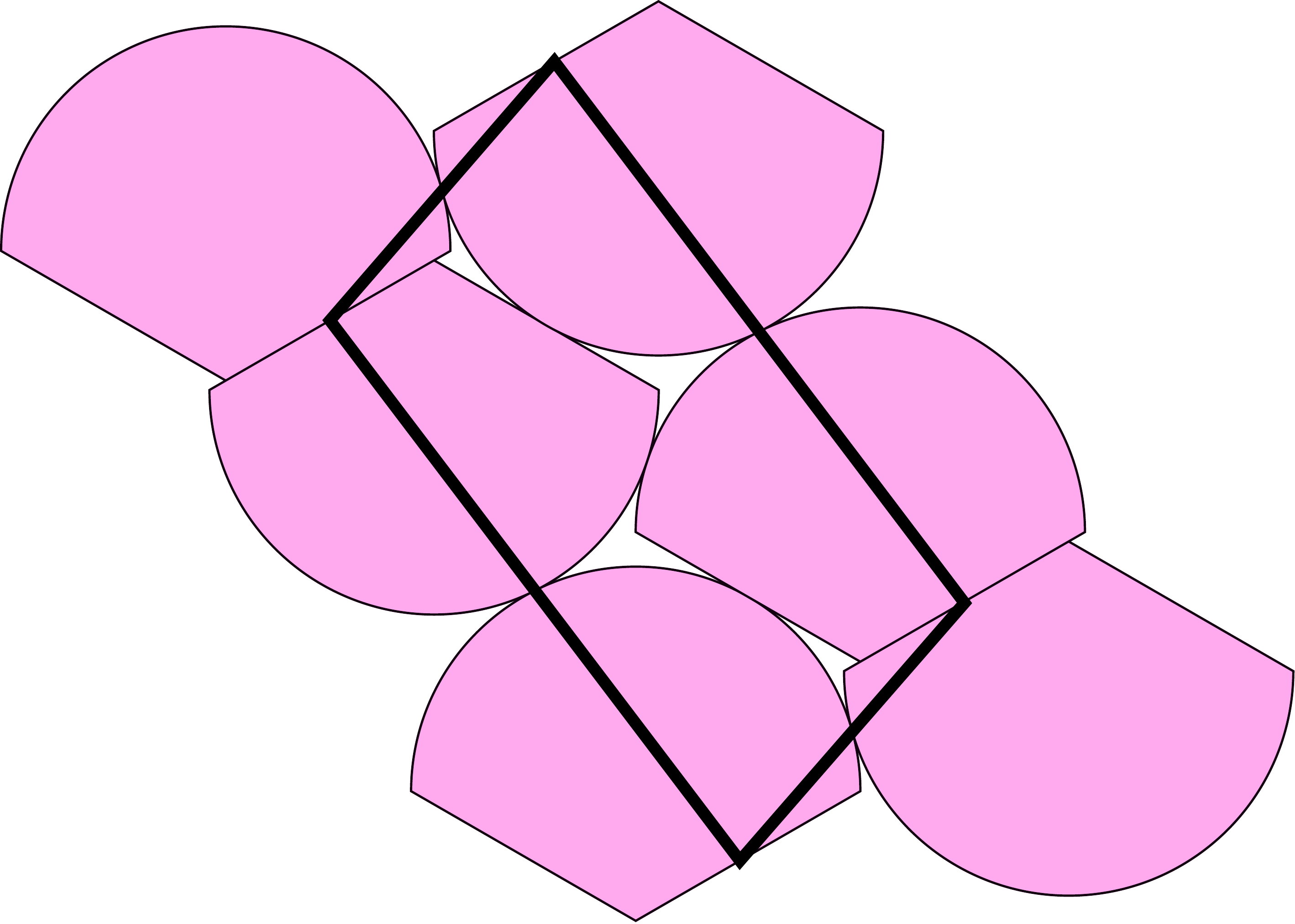}}
            \hfill
            \subfloat[]{\label{fig:T2_ex}
                \includegraphics[height=0.225\textheight]{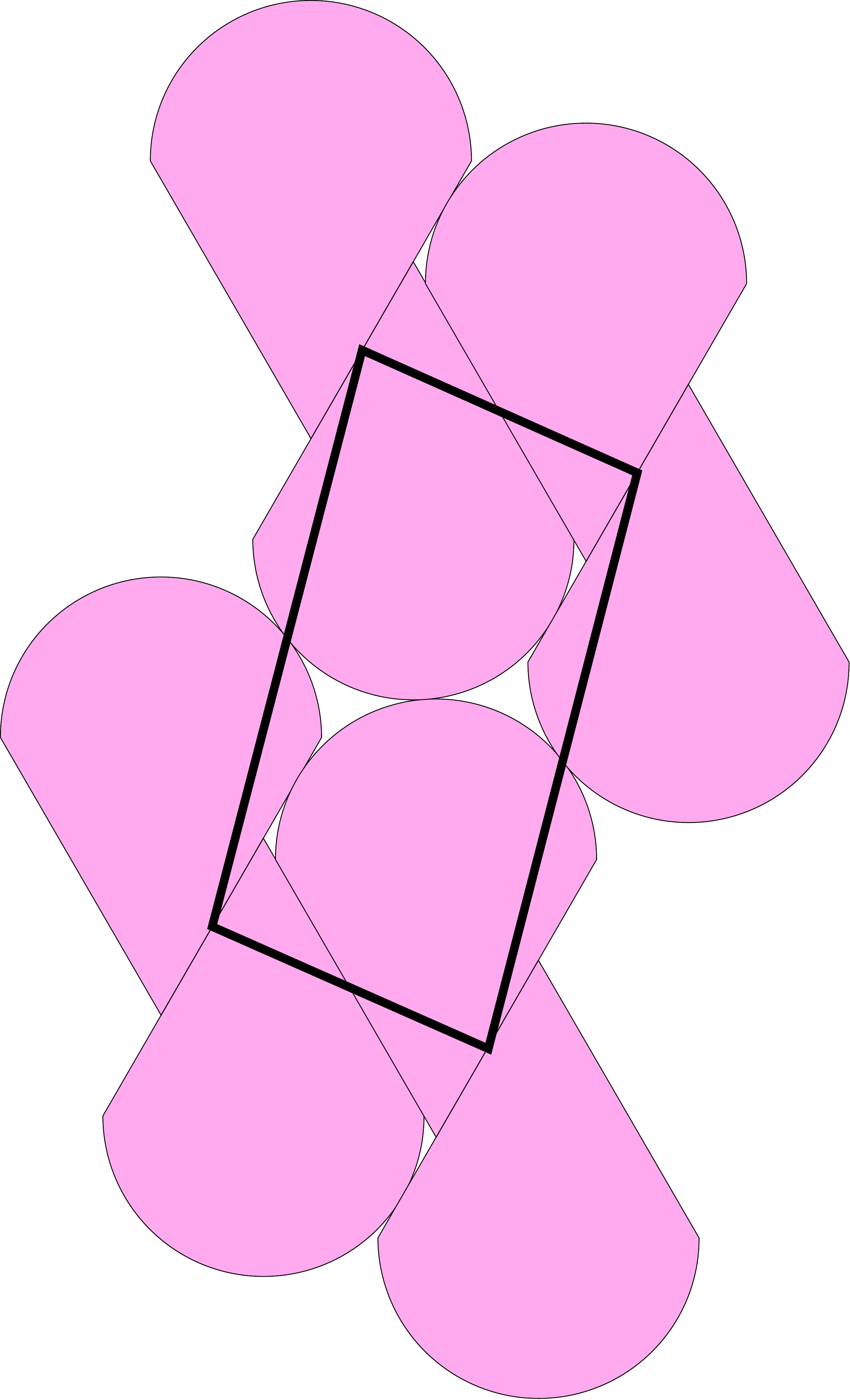}}
                
            \caption{Fundamental bases for (a) a type 1 ($\theta_{ICC}=120^{\circ}$) ice cream cone packing and (b) a type 2 ($\theta_{ICC}=60^{\circ}$) ice cream cone packing.}
            \label{fig:DensestPackings}
    \end{figure}

\begin{figure}[t]
        \centering
        \includegraphics[width=0.4\textwidth]{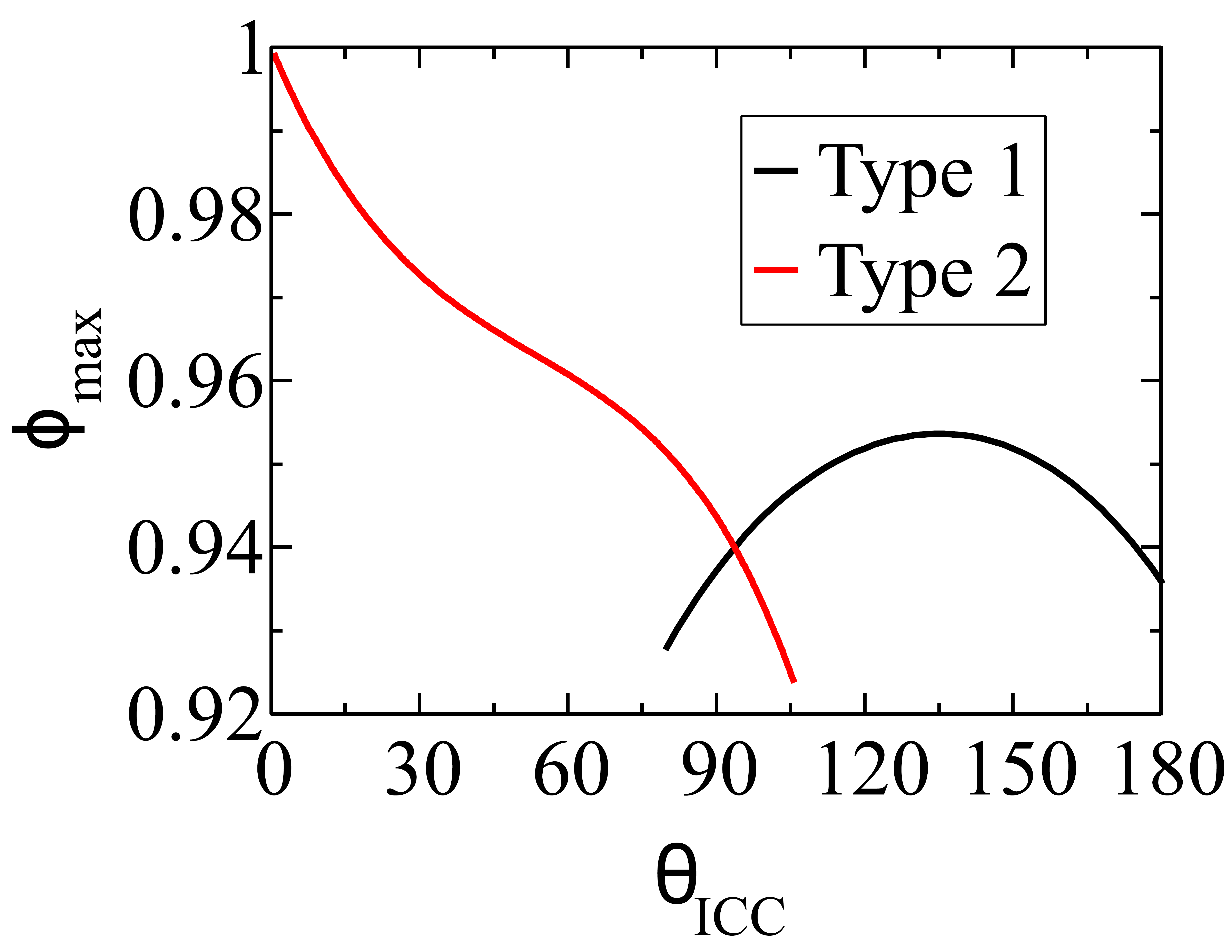}
        \caption{Putative maximum packing fraction $\phi_{max}$ as a function of $\theta_{ICC}$ for both types of dense ice cream cone packings.}
        \label{fig:ICC_DenseFunction}
\end{figure}

\begin{figure*}[t!]
        \centering
        \subfloat[]{\label{fig:504Romb}
        \includegraphics[width = 0.35\textwidth]{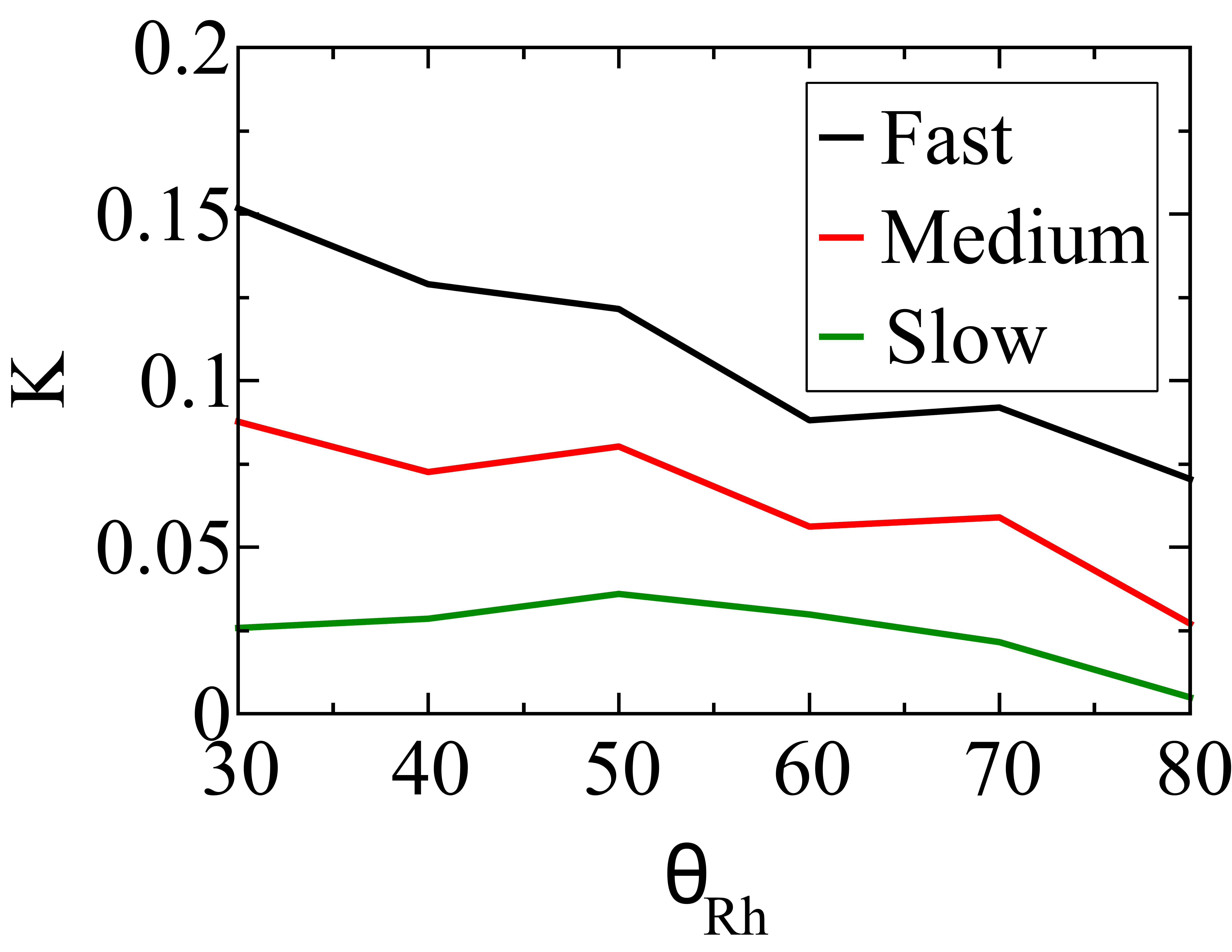}
        }
        \subfloat[]{\label{fig:504OST}
        \includegraphics[width = 0.35\textwidth]{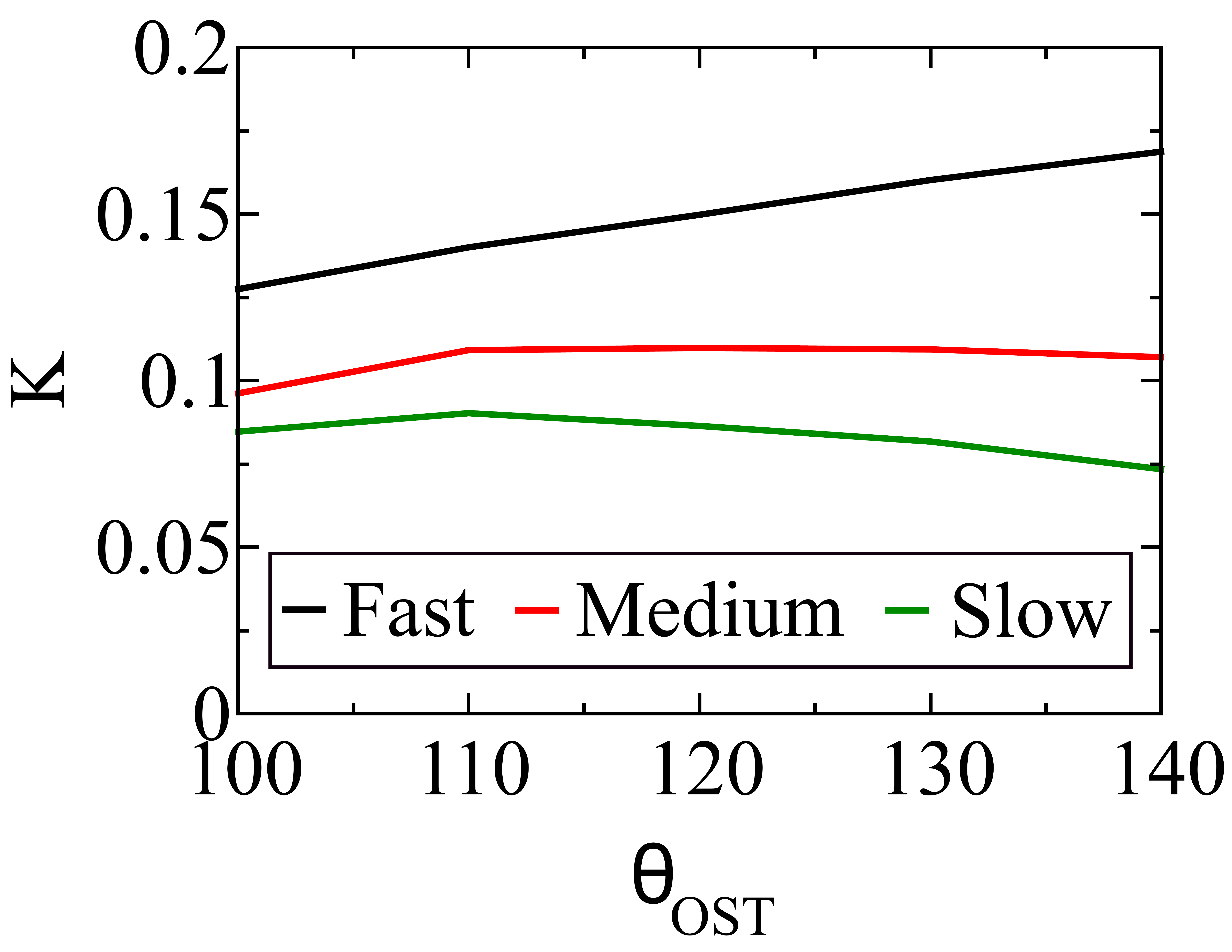}
        }
        
        \subfloat[]{\label{fig:504CT}
        \includegraphics[width = 0.35\textwidth]{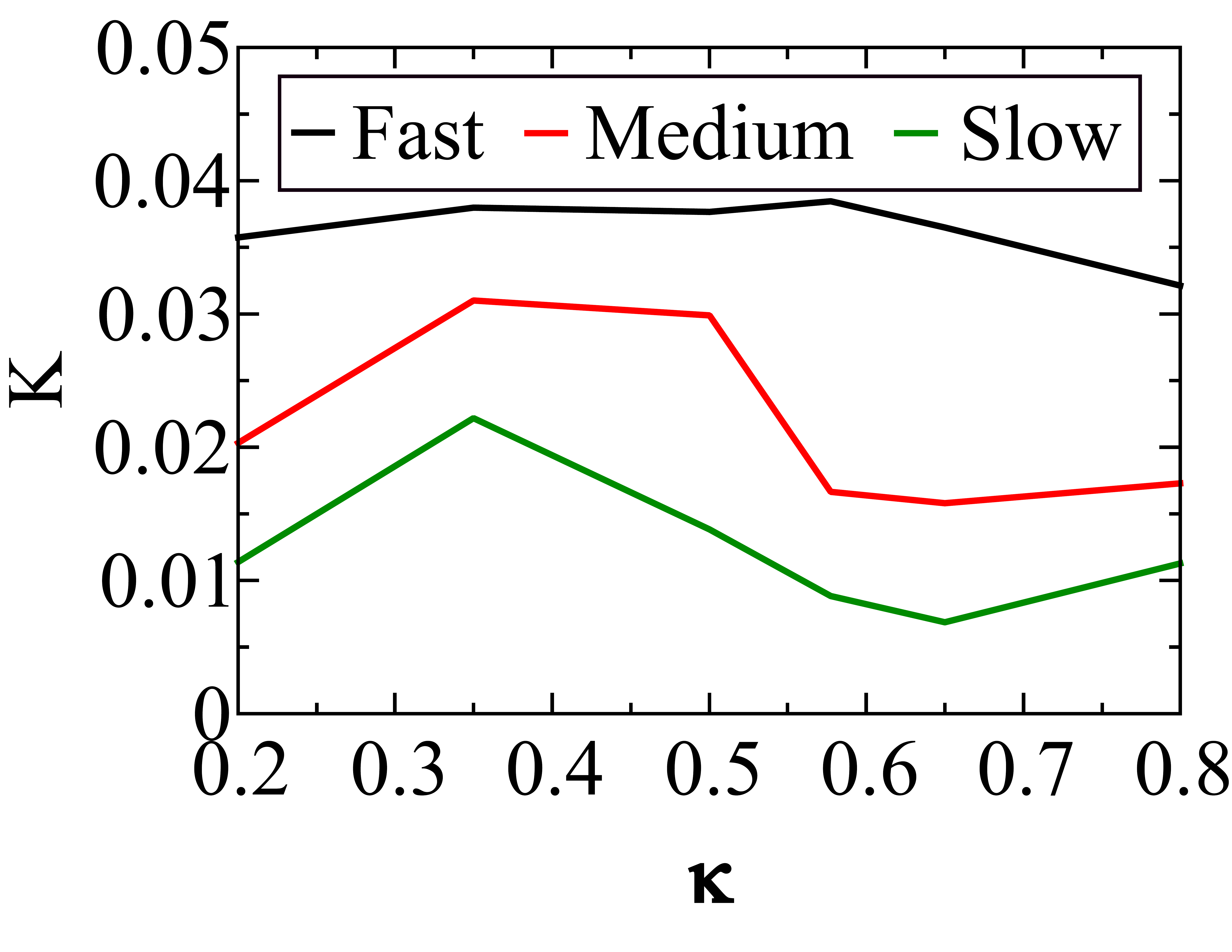}
        }
        \subfloat[]{\label{fig:504Lens}
        \includegraphics[width = 0.35\textwidth]{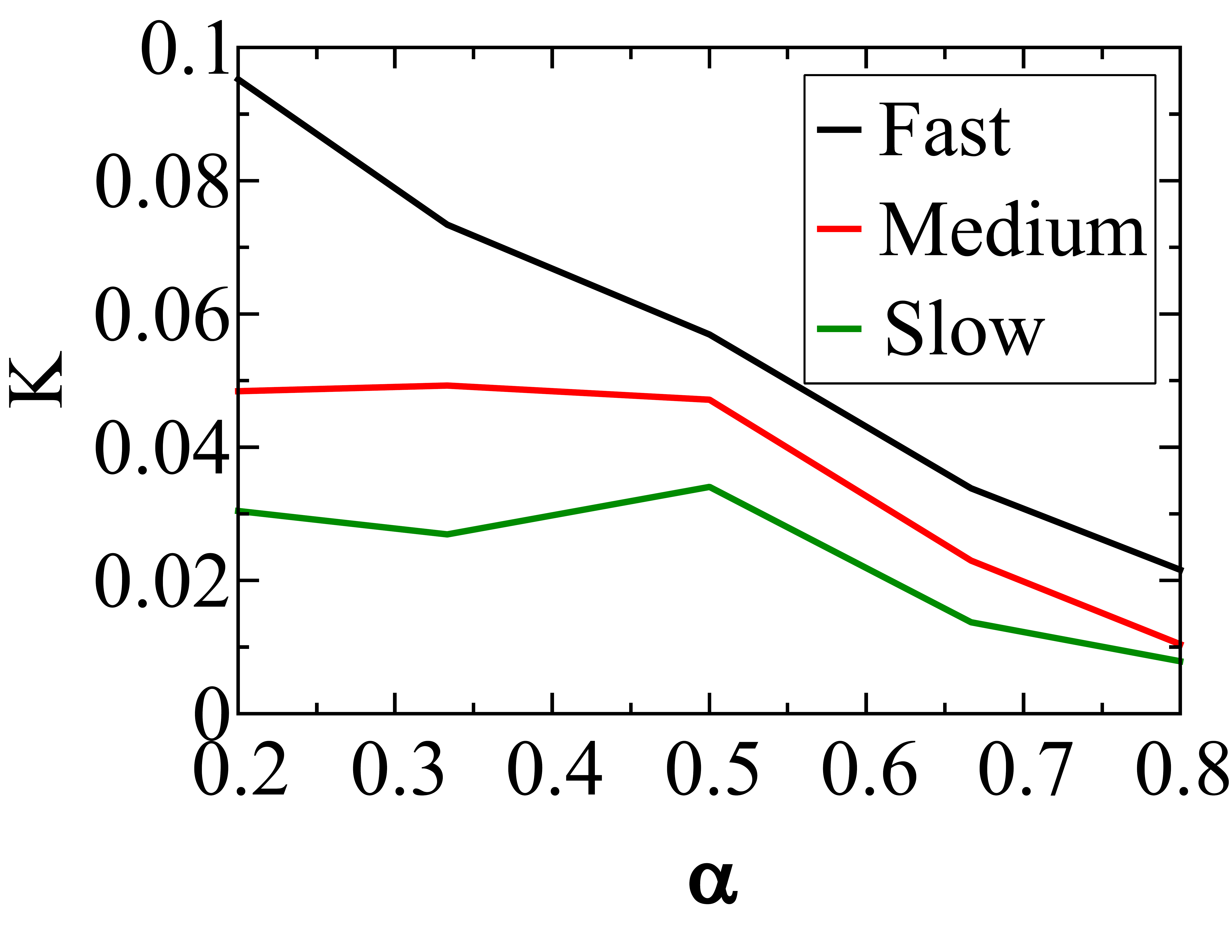}
        }
        
        \subfloat[]{\label{fig:504ICC}
        \includegraphics[width = 0.35\textwidth]{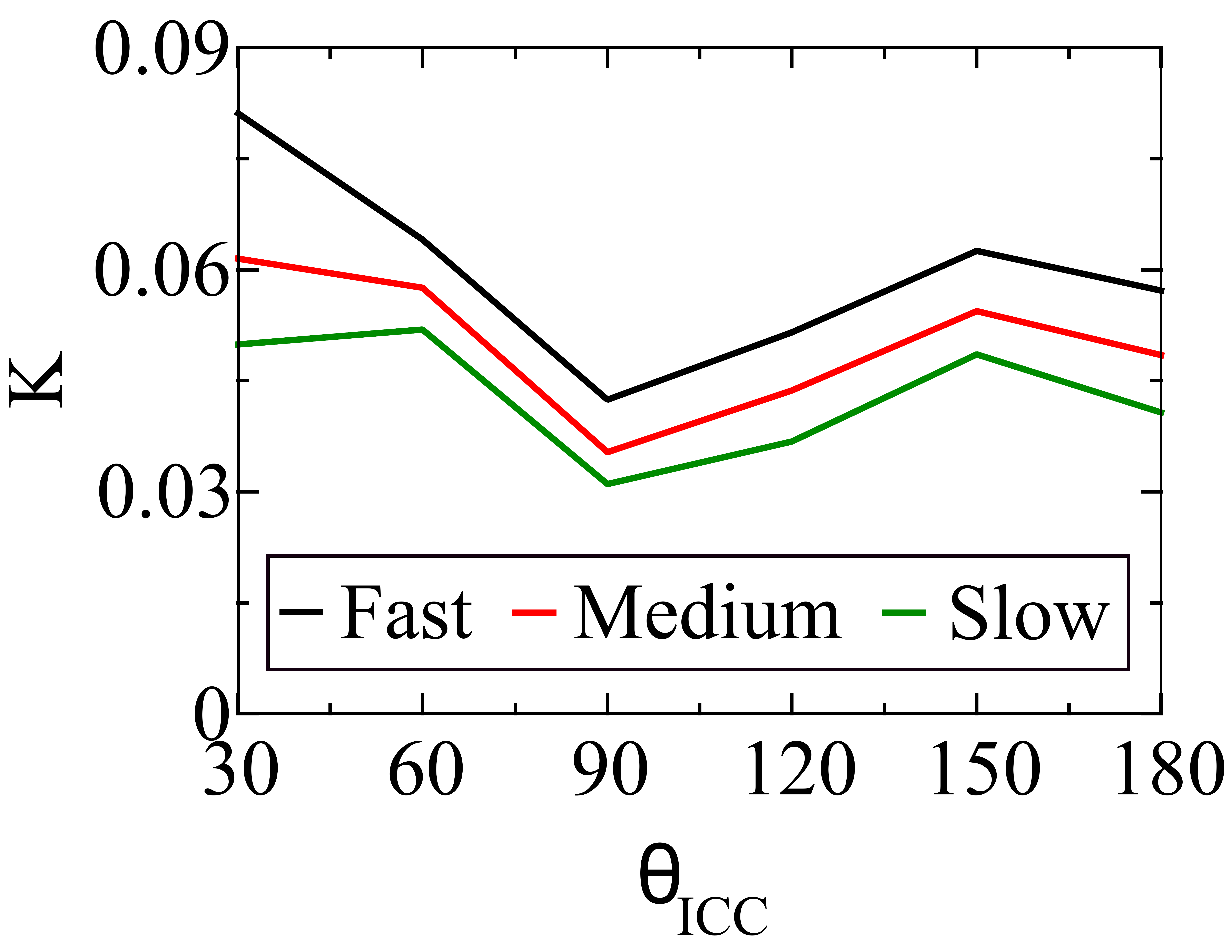}
        }
        \caption{Kinetic frustration index $K$ for (a) rhombi as a function of $\theta_{Rh}$, (b) OST as a function of $\theta_{OST}$, (c) CT as a function of $\kappa$, (d) lenses as a function of $\alpha$, and (e) ICC as a function of $\theta_{ICC}$.}
        \label{fig:K_data}
\end{figure*} 

\paragraph{Ice Cream Cones}\label{Sec:DenseICC}
The putative densest packings of ICCs are consistent with the conjecture that noncentrally symmetric objects have their maximum densities realized in a two-particle basis double lattice packing (defined in section \ref{sec:Packings}) \cite{Kuperberg_x2}.
We have found two distinct packing behaviors, which we name ``type 1" and ``type 2" packings, both of which are consistent with the theorem due to Kuperberg and Kuperberg\cite{Kuperberg_x2} stating that each two-dimensional convex body admits a double lattice packing with $\phi > \sqrt{3}/2$.
Type 1 packings are densest for ICCs with large $\theta_{ICC}$, and type 2 packings are densest for ICCs with small $\theta_{ICC}$.
The fundamental bases of both packing types are shown in Figure \ref{fig:DensestPackings}.
Figure \ref{fig:ICC_DenseFunction} shows $\phi_{max}$ for all values of $\theta_{ICC}$.
In type 1 packings, $\phi_{max}$ does not vary monotonically, and in the semicircle limit ($\theta_{ICC}\rightarrow180^{\circ}$), $\phi = 0.935931$, a value that is consistent with the exact result given in ref \citenum{Semicirc_Dense}, namely
\begin{equation}
\phi = \frac{\pi}{\sqrt{3}+5\textrm{tan}\frac{\pi}{10}}=0.9359311\dots
\end{equation}
This level of agreement with the theoretical prediciton is a testament to the efficacy of the ASC scheme to produce the densest jammed packing.
Moreover, in type 2 packings $\phi_{max}$ decreases monotonically with $\theta_{ICC}$ and reaches unity in the limit of $\theta_{ICC}\rightarrow0$.

\section{Results and Discussion}\label{sec:Results}

Here, we use the stochastic solution to the ASC scheme and the three compression schedules described in section \ref{sec:ASCScheme} to generate dense monodisperse packings of the shapes described in section \ref{sec:ShapeDefns}.
We then characterize the kinetic frustration, degree of short- and long-range order via the spectral density, and contact networks in these packings as a function of particle shape and compression rate.

\subsection{Kinetic Frustration}\label{sec:Kinetics}

The kinetic frustration index $K$ as a function of the particle shape and compression rate is given in Figure \ref{fig:K_data}, which shows that $K$ increases as the compression rate increases, regardless of particle shape.
Three shape properties that have a demonstrable effect on $K$ are rotational symmetry, curvature, and asphericity $\gamma$.

\begin{figure*}[t!]
        \centering
        \subfloat[]{\label{fig:100OST_F2}
        \includegraphics[width = 0.95\textwidth]{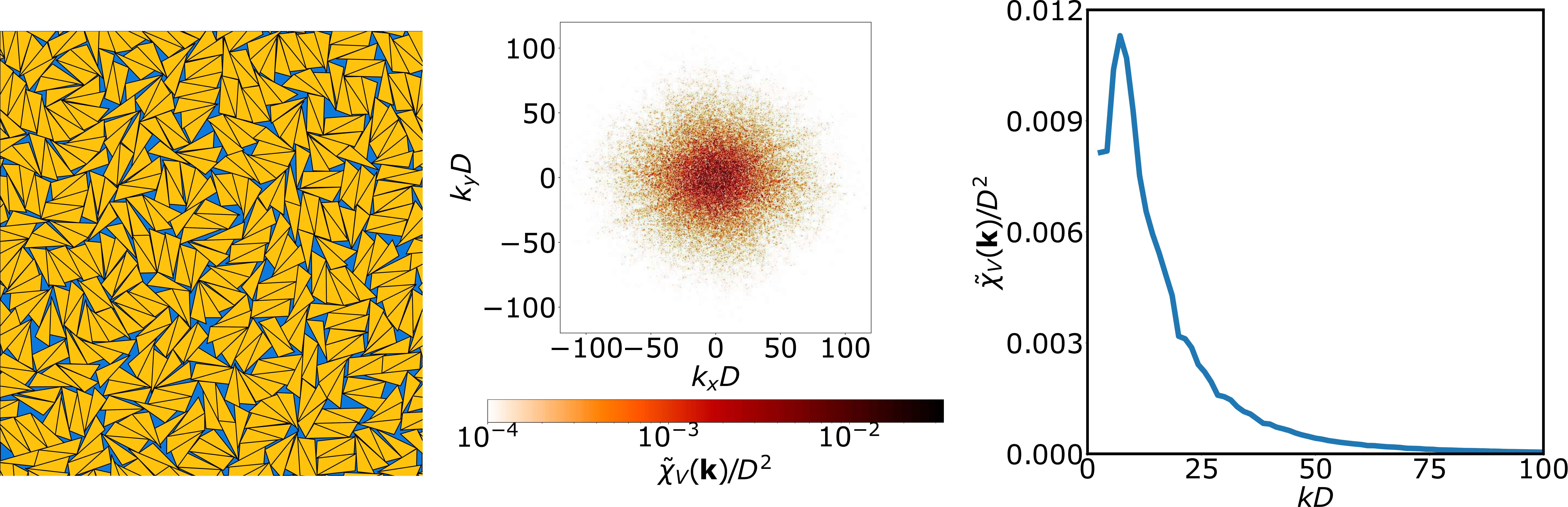}
        }
        
        \subfloat[]{\label{fig:140OST_S0}
        \includegraphics[width = 0.95\textwidth]{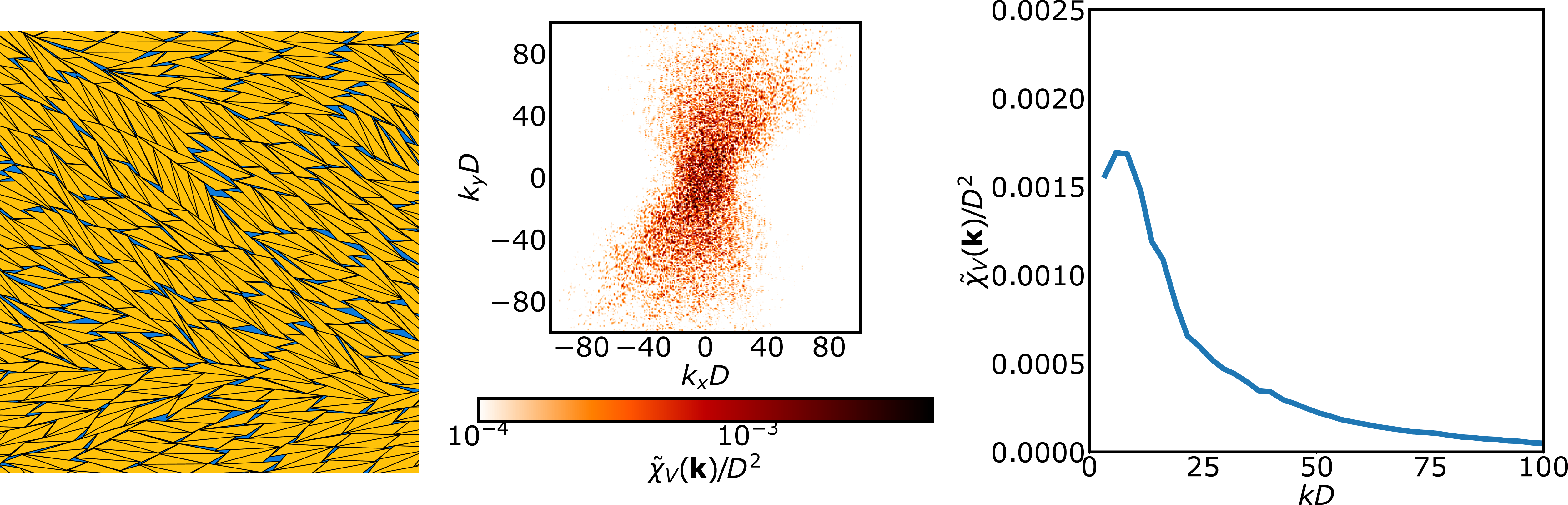}
        }
        \caption{Example representative configurations (left), corresponding two-dimensional (center), and angular averaged spectral densities (right) $\spD{\vect{k}}/D^2$ vs dimensionless wavenumber $kD$ from OST packings: (a) $\theta_{OST}=100^{\circ}$ OST compressed using the fast schedule and (b) $\theta_{OST}=140^{\circ}$ OST compressed using the slow schedule, where $D$ is the circumradius of the particle.}
        \label{fig:OST_Spect}
\end{figure*}

\subsubsection{Symmetry}\label{sec:Symmetry}
The packings of OST, which are the least symmetric particles studied here, tend to have a larger $K$ than other packings (see Figure \ref{fig:K_data}b).
This can be directly contrasted with the results for the packings of rhombi, which have 2-fold rotational symmetry (see Figure \ref{fig:K_data}a), and tend to have much lower values of $K$.
Comparison of CT and lens results (see Figures \ref{fig:K_data}c and \ref{fig:K_data}d, respectively) indicates curved shapes with similar $\gamma$ have similar $K$, despite having different degrees of rotational symmetry.
Thus, increased rotational symmetry reduces the kinetic frustration of polygonal particle shapes but does not significantly affect curved particle shapes.

\subsubsection{Curvature}\label{sec:Curvature}
Directly comparing packings of curved (e.g., Figure \ref{fig:K_data}d) and faceted shapes (e.g., Figure \ref{fig:K_data}a) shows that for shapes with similar $\gamma$ $K$ is lower in packings of curved shapes.
This is also evident in CT packings (see Figure \ref{fig:K_data}c) where smaller $\kappa$ (particle curvature) tends to result in larger $K$. 
This is not observed in slowly compressed CT packings in the $\kappa\rightarrow0$ limit due to the emergence of 6-fold orientational order (see section \ref{sec:SpecDens}).
In ICC packings, $K$ increases faster as $\theta_{ICC}$ deviates from 90$^{\circ}$ toward 0$^{\circ}$ than it does as $\theta_{ICC}$ deviates from 90$^{\circ}$ toward 180$^{\circ}$ (see Figure \ref{fig:K_data}e).
This behavior is attributed to the fraction of the shape perimeter that is curved going to zero as $\theta_{ICC}$ goes to 0$^{\circ}$.

\begin{figure*}[t!]
        \centering
        \subfloat[]{\label{fig:30Rh_F0}
        \includegraphics[width = 0.95\textwidth]{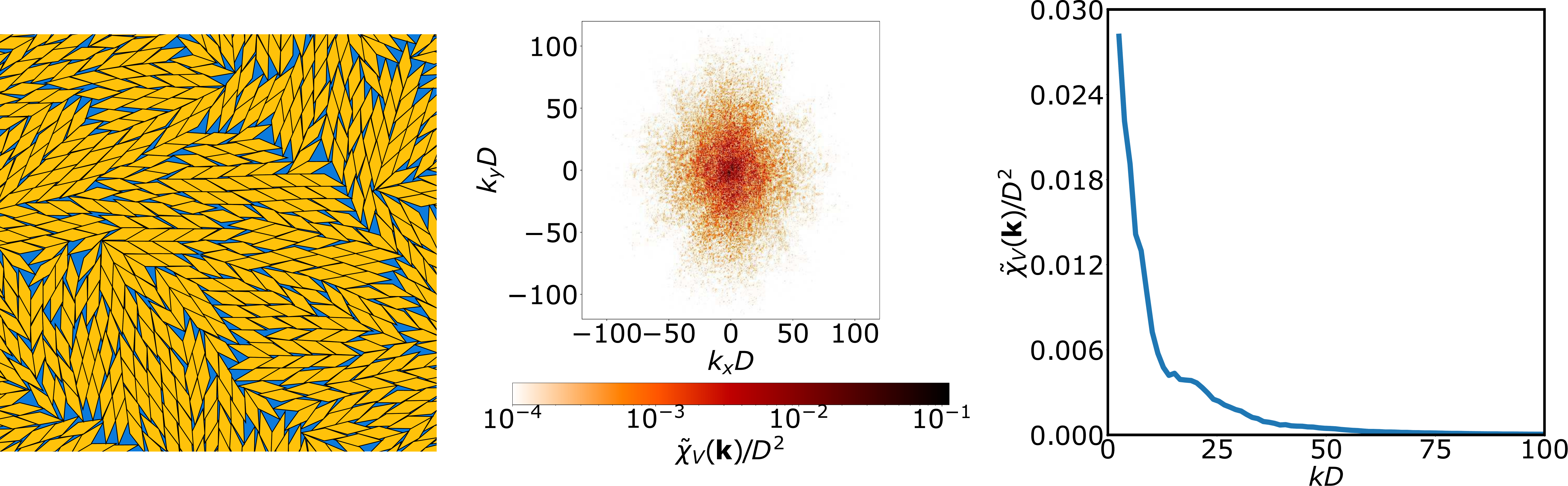}
        }
        
        \subfloat[]{\label{fig:80Rh_F0}
        \includegraphics[width = 0.95\textwidth]{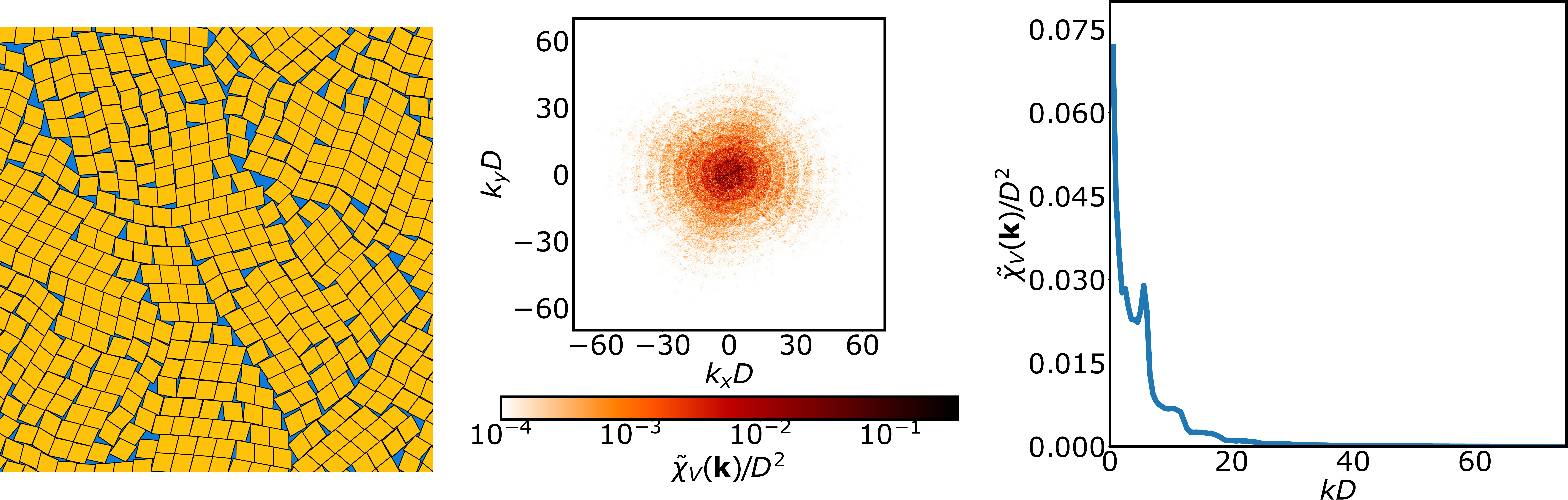}
        }

        \subfloat[]{\label{fig:80Rh_S0}
        \includegraphics[width = 0.95\textwidth]{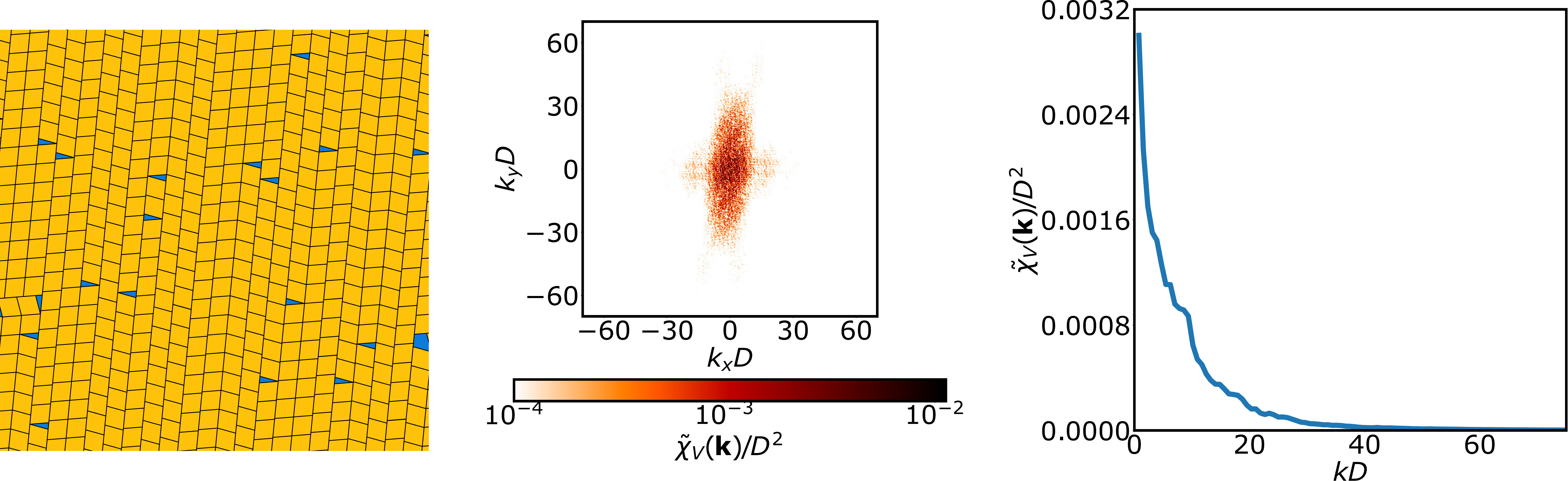}
        }
        \caption{Example representative configurations (left), corresponding two-dimensional (center), and angular averaged spectral densities (right) spectral densities $\spD{\vect{k}}/D^2$ vs dimensionless wavenumber $kD$ from rhombus packings: (a) $\theta_{Rh} = 30^{\circ}$ rhombi compressed using the fast schedule, (b) $\theta_{Rh} = 80^{\circ}$ rhombi compressed using the fast schedule, and (c) $\theta_{Rh} = 80^{\circ}$ rhombi compressed using the slow schedule, where $D$ is the circumradius of the particle.}
        \label{fig:Rh_Spect}
\end{figure*} 

\subsubsection{Asphericity}\label{sec:Asphere}
Rhombus and lens packings show an increase in $K$ as $\gamma$ increases, indicating that greater asphericity results in greater kinetic frustration.
This trend can also be observed in Figure \ref{fig:K_data}e where both $\gamma$ and $K$ increase as $\theta_{ICC}$ deviates from 90$^{\circ}$.
Rapidly compressed OST packings also exhibit this behavior (see Figure. \ref{fig:K_data}b), but more slowly compressed OST packings do not due to the emergence of 2-fold orientational order (see section \ref{sec:SpecDens}).

\begin{figure*}[t!]
        \centering
        \subfloat[]{\label{fig:720CT_S0}
        \includegraphics[width = 0.95\textwidth]{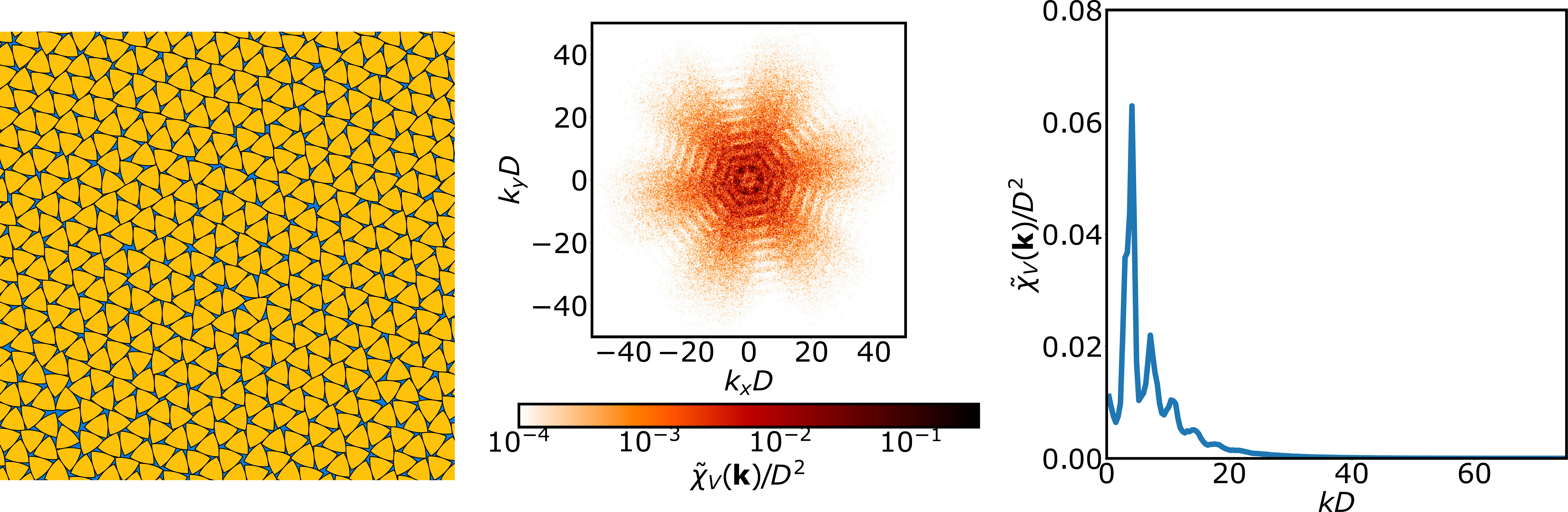}
        }
        
        \subfloat[]{\label{fig:12CT_F0}
        \includegraphics[width = 0.95\textwidth]{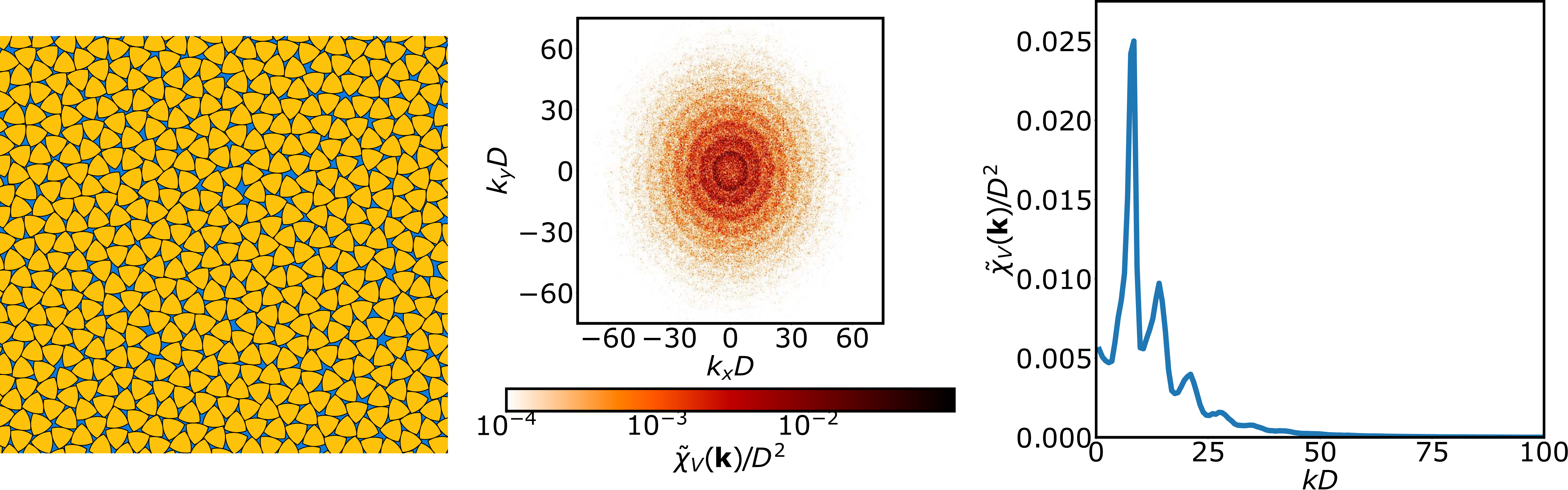}
        }
        
        \subfloat[]{\label{fig:RTCT_S0}
        \includegraphics[width = 0.95\textwidth]{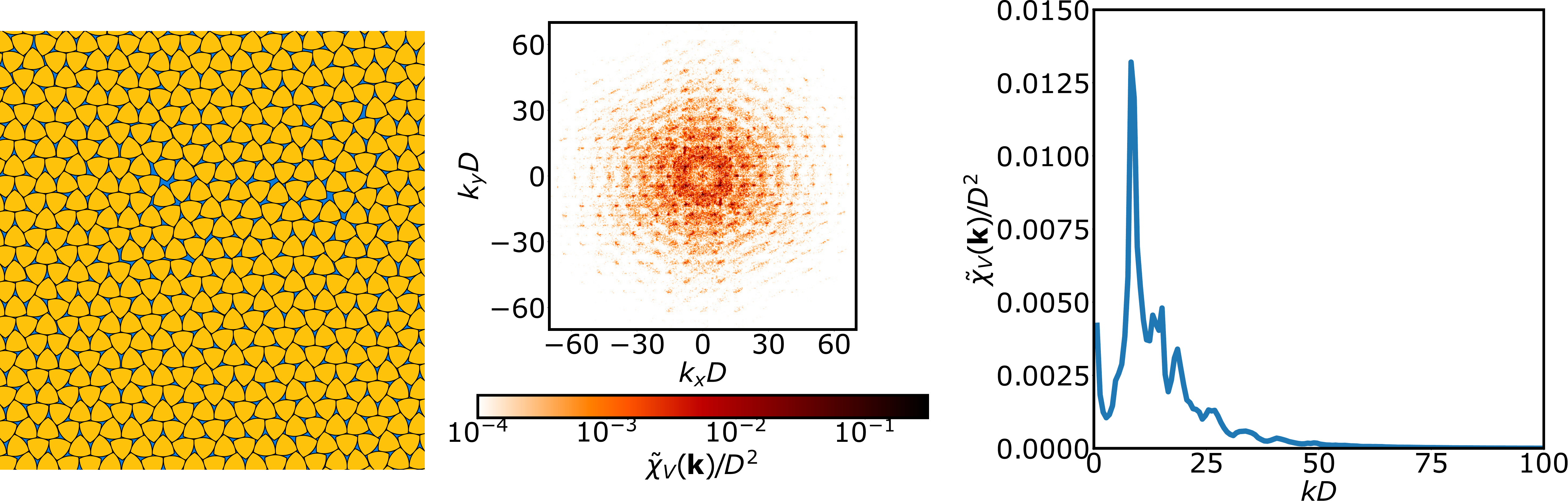}
        }
        \caption{Example representative configurations (left), corresponding two-dimensional (center), and angular averaged spectral densities (right) densities $\spD{\vect{k}}/D^2$ vs dimensionless wavenumber $kD$ from CT packings: (a) $\kappa=0.35$ CT compressed using the slow compression schedule, (b) $\kappa=0.5$ CT compressed using the fast compression schedule, and (c) Reuleaux triangles ($\kappa=1/\sqrt{3}$) compressed using the slow compression schedule, where $D$ is the circumradius of the particle.}
        \label{fig:CT_Spect}
\end{figure*}

\subsection{Spectral Density}\label{sec:SpecDens}
Through visualization of the spectral density we characterize the short- and long-range translational and orientational order in the packings produced through the ASC scheme as a function of compression rate and particle shape.
Here, we examine both two-dimensional and angular averaged $\spD{\vect{k}}$.

\begin{figure*}[t!]
        
        \centering
        \subfloat[]{\label{fig:180ICC_F0}
        \includegraphics[width = 0.95\textwidth]{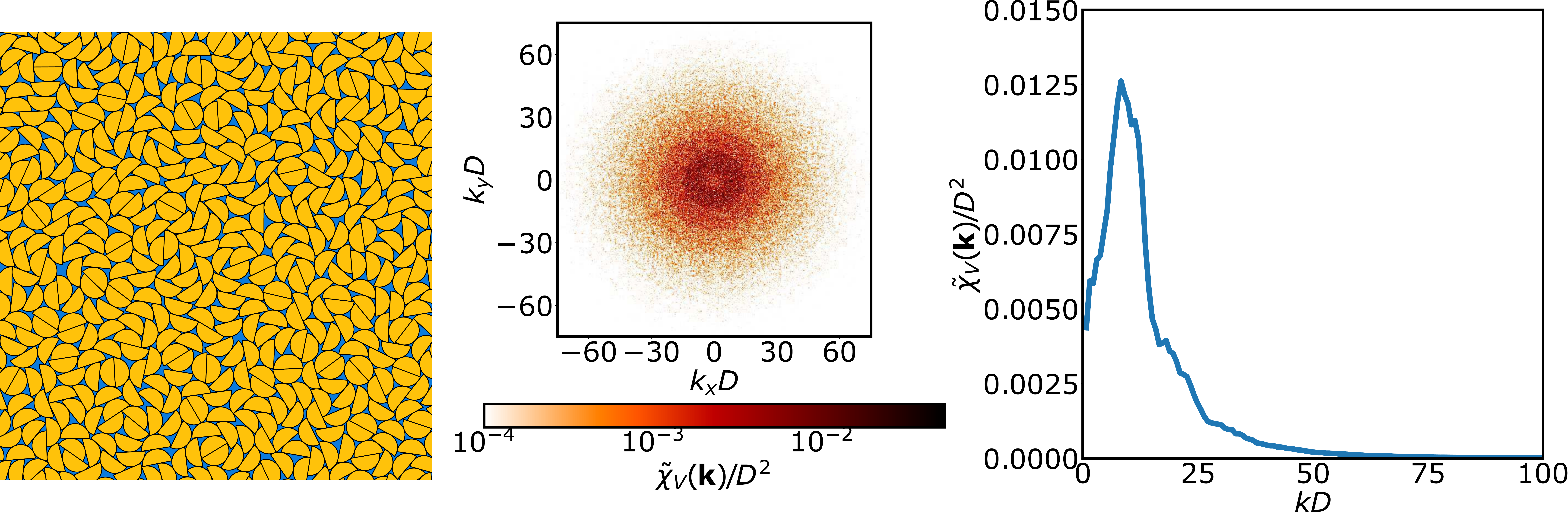}
        }
        
        \subfloat[]{\label{fig:90ICC_M0}
        \includegraphics[width = 0.95\textwidth]{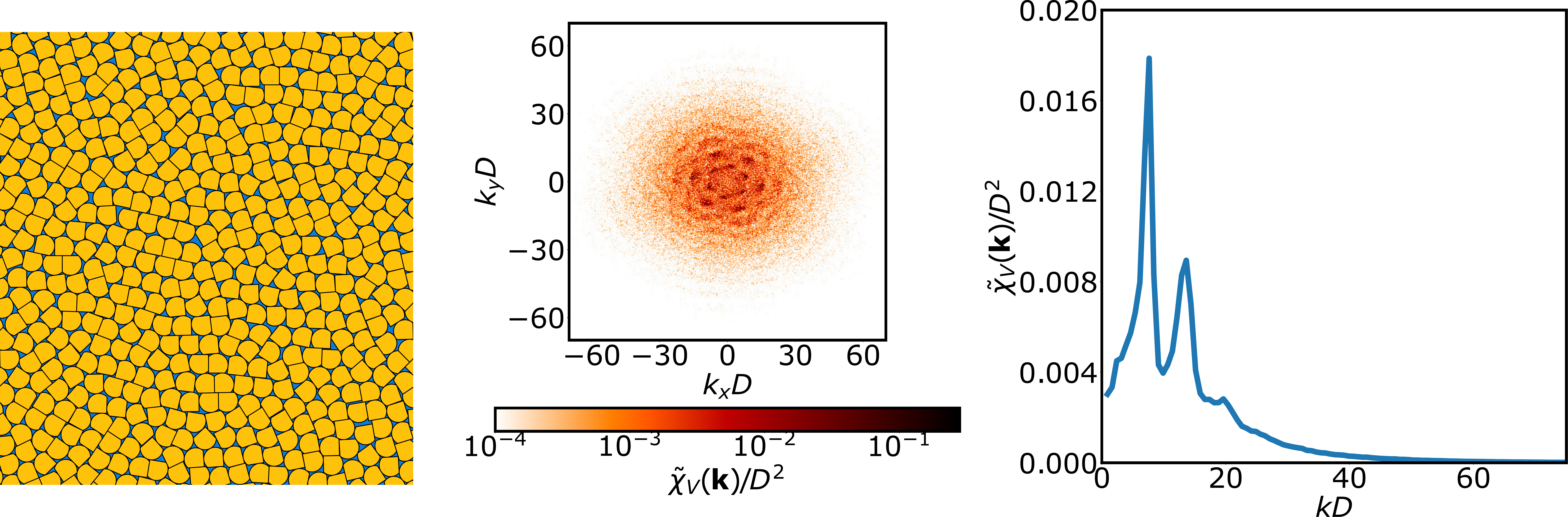}
        }
        
        \subfloat[]{\label{fig:30ICC_F0}
        \includegraphics[width = 0.95\textwidth]{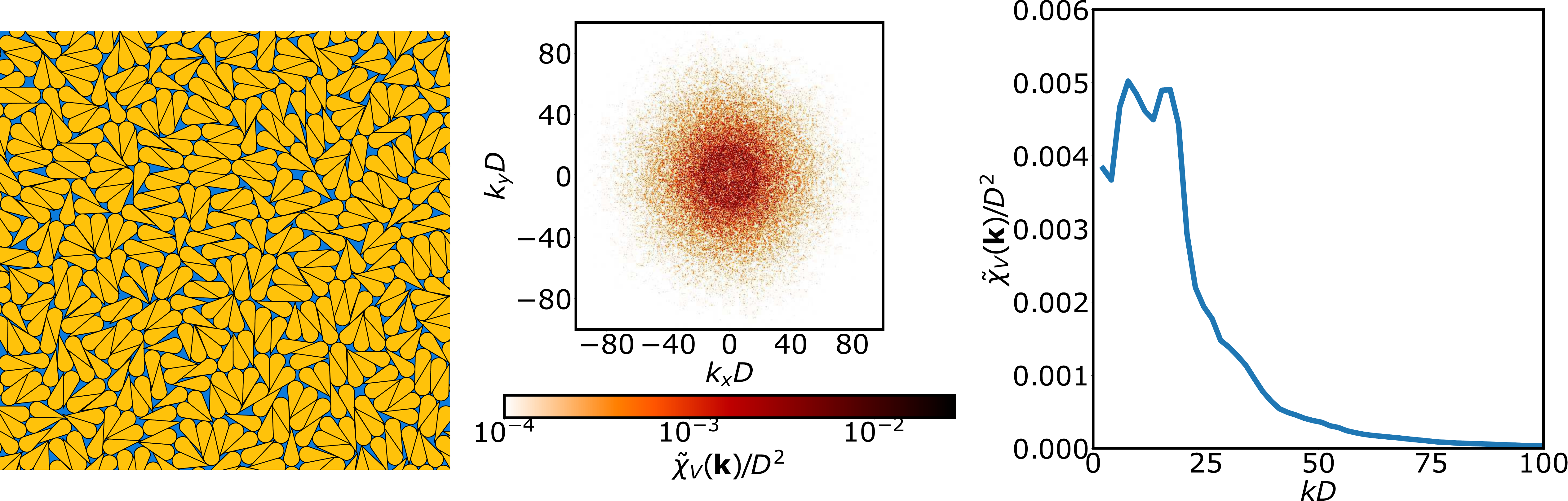}
        }
        \caption{Example representative configurations (left), corresponding two-dimensional (center), and angular averaged spectral densities (right) spectral densities $\spD{\vect{k}}/D^2$ vs dimensionless wavenumber $kD$ from ICC packings: (a) $\theta_{ICC}=180^{\circ}$ ICC compressed using the fast compression schedule, (b) $\theta_{ICC}=90^{\circ}$ ICC compressed using the medium compression schedule, and (c) $\theta_{ICC}=30^{\circ}$ ICC compressed using the fast compression schedule, where $D$ is the circumradius of the particle.}
        \label{fig:ICC_Spect}
\end{figure*} 

There does not appear to be significant short-range translational order in any of the OST packings, shown by the lack of distinct peaks in the angular-averaged spectral densities in Figure \ref{fig:OST_Spect}.
Broad peaks close to the origin indicate that there are large-scale density fluctuations in these OST packings, which diminish in intensity as $kD$ increases.
Short-range nematic (2-fold) orientational order in OST packings increases as $\theta_{OST}$ increases and as the compression rate decreases.
This behavior is exemplified by the non-radially symmetric two-dimensional spectral density in Figure \ref{fig:OST_Spect}b, which is lost upon angular averaging.

\begin{figure*}[t!]
        \centering
        \subfloat[]{\label{fig:45Lens_S0}
        \includegraphics[width = 0.95\textwidth]{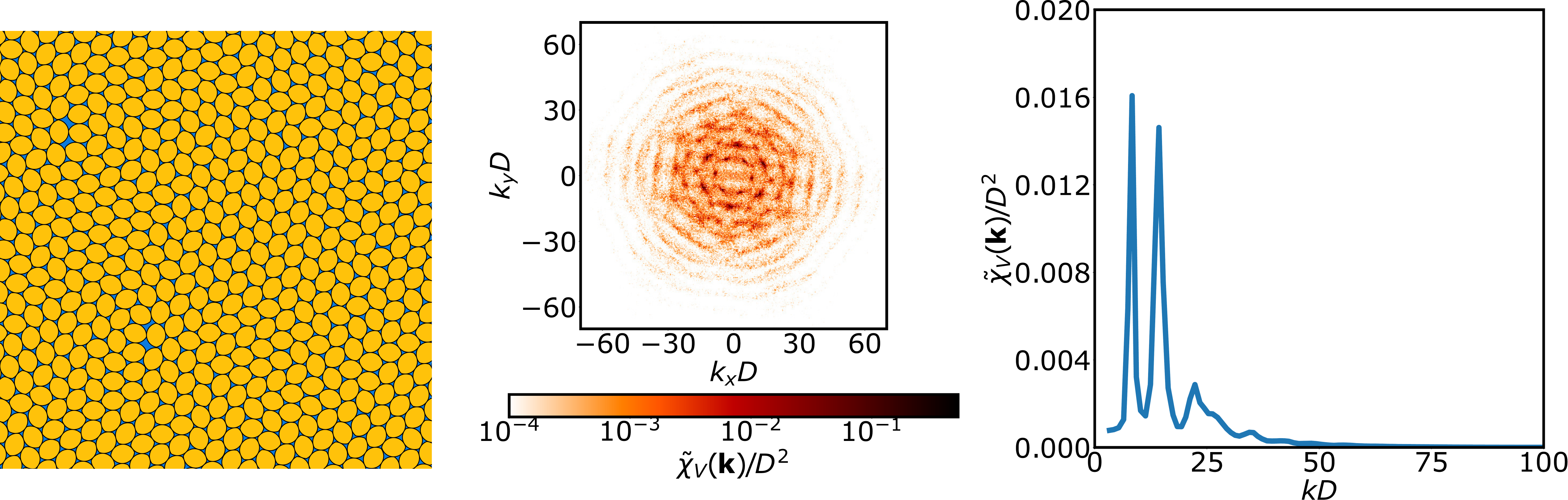}
        }
        
        \subfloat[]{\label{fig:23Lens_Med}
        \includegraphics[width = 0.95\textwidth]{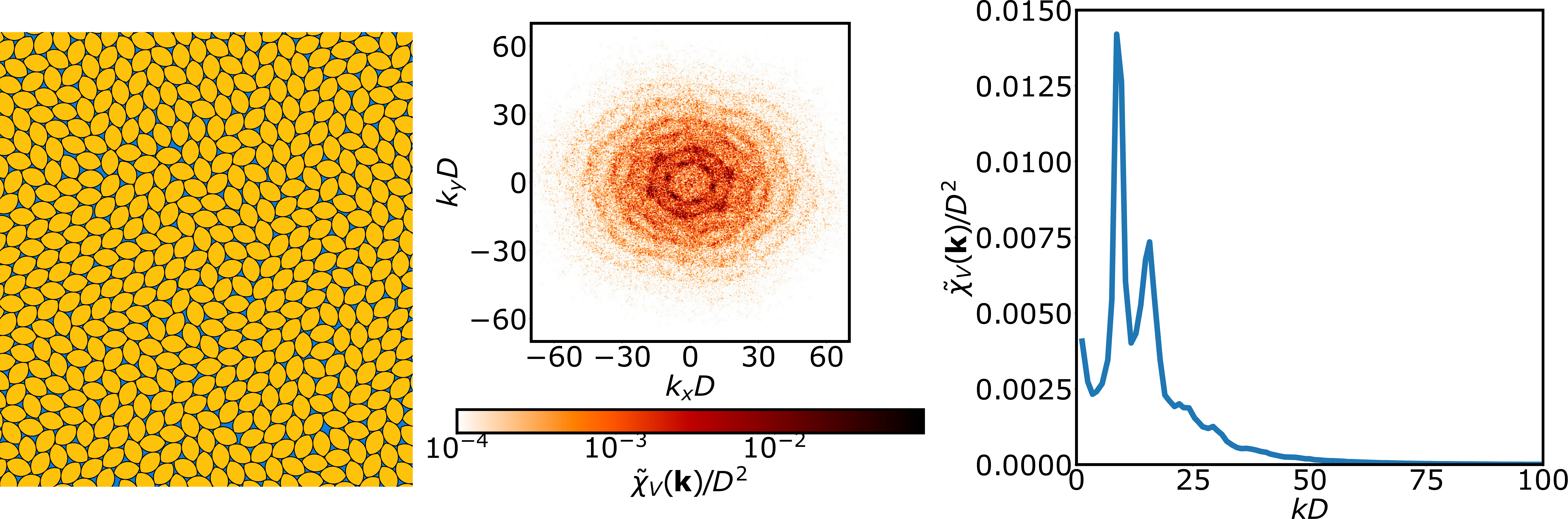}
        }
        
        \subfloat[]{\label{fig:15Lens_S0}
        \includegraphics[width = 0.95\textwidth]{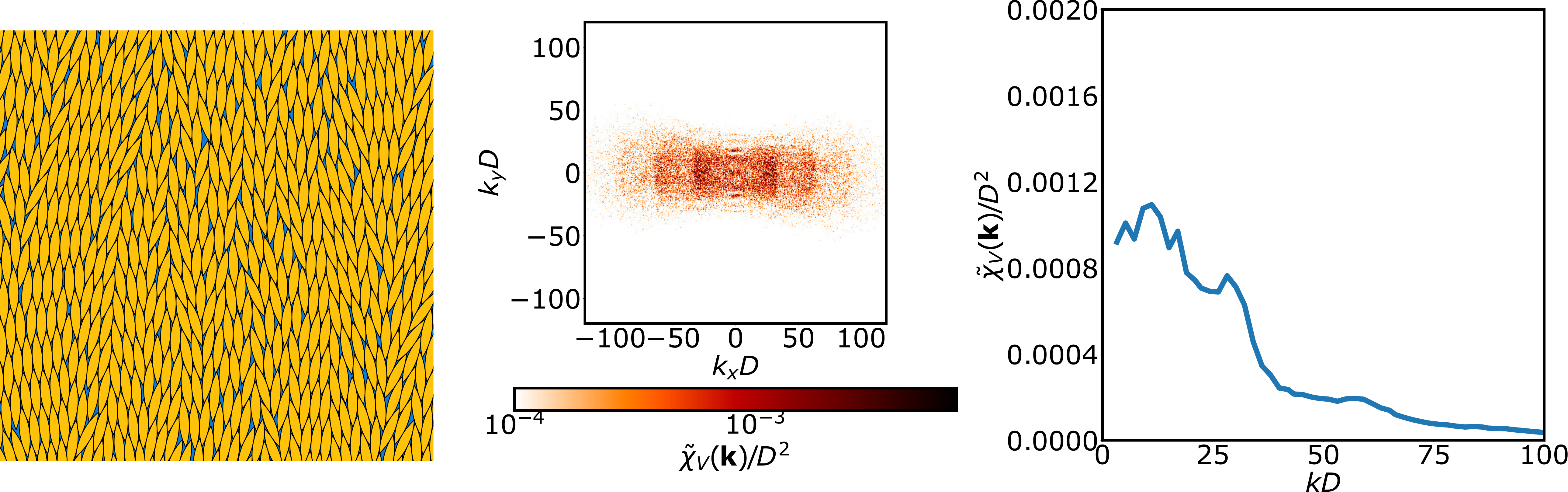}
        }
        \caption{Example representative configurations (left), corresponding two-dimensional (center), and angular averaged spectral densities (right) spectral densities $\spD{\vect{k}}/D^2$ vs dimensionless wavenumber $kD$ from lens packings: (a) $\alpha = 0.8$ lenses compressed using the slow compression schedule, (b) $\alpha = 0.666...$ lenses compressed using the medium compression schedule, and (c) $\alpha = 0.2$ lenses compressed using the slow compression schedule, where $D$ is the circumradius of the particle.}
        \label{fig:Lens_Spect}
\end{figure*} 

In rhombus packings, the degree of short-range translational order increases with the value of $\theta_{Rh}$.
The two-dimensional spectral density in the center panel of Figure \ref{fig:Rh_Spect}c indicates short- and long-ranged translational and nematic orientational order.
However, the large fluctuations in the distances between defects results in an angular-averaged spectral density that appears to lack translational order (see right panel of Figure \ref{fig:Rh_Spect}c).
Otherwise, as $\theta_{Rh}$ increases, the spectral densities begin to exhibit peaks.
The high scattering intensity near the origin in the right panels of Figure \ref{fig:Rh_Spect} indicates large-scale density fluctuations that decrease in intensity as $kD$ increases.
In the right panel of Figure \ref{fig:Rh_Spect}b there is a peak at $kD\sim6$, which corresponds to density fluctuations on the order of the size of a single rhombus.
Slowly compressed packings of rhombi exhibit short-range nematic orientational order (see center panels of Figures \ref{fig:Rh_Spect}a and \ref{fig:Rh_Spect}c), which is lost up angular averaging. 
As the compression rate increases, the packings become more isotropic (see center panel of Figure \ref{fig:Rh_Spect}b).
We also observe that these packings are polycrystalline, and slowly compressed rhombi with small $\gamma$ tend to exhibit point-like defects, while lager compression rates and $\gamma$ result in grain boundary-like defects.

All CT packings exhibit a significant degree of short-range translational order, shown by the peaks in the right panels of Figure \ref{fig:CT_Spect}.
The first peak in the right panels of Figure \ref{fig:CT_Spect} corresponds to a length scale associated with the distance between neighboring voids between the particles.
In right panel of (c) of Figure \ref{fig:CT_Spect} the split second peak corresponds to length scales associated with the height of the CT and the side length of the CT.
Short-range 6-fold orientational order that is lost upon angular averaging emerges when the packings are compressed more slowly (see center panels of Figures \ref{fig:CT_Spect}a and \ref{fig:CT_spect}c), while fast compressions can result in isotropic packings (see center panel of Figure \ref{fig:CT_Spect}b).
As $\kappa$ increases, the six-fold orientational order increases in range, shown by the sharpening of the Bragg peaks in the center panel of Fig. \ref{fig:CT_Spect}c.

\begin{figure}[t!]
    \includegraphics[width=0.46\textwidth]{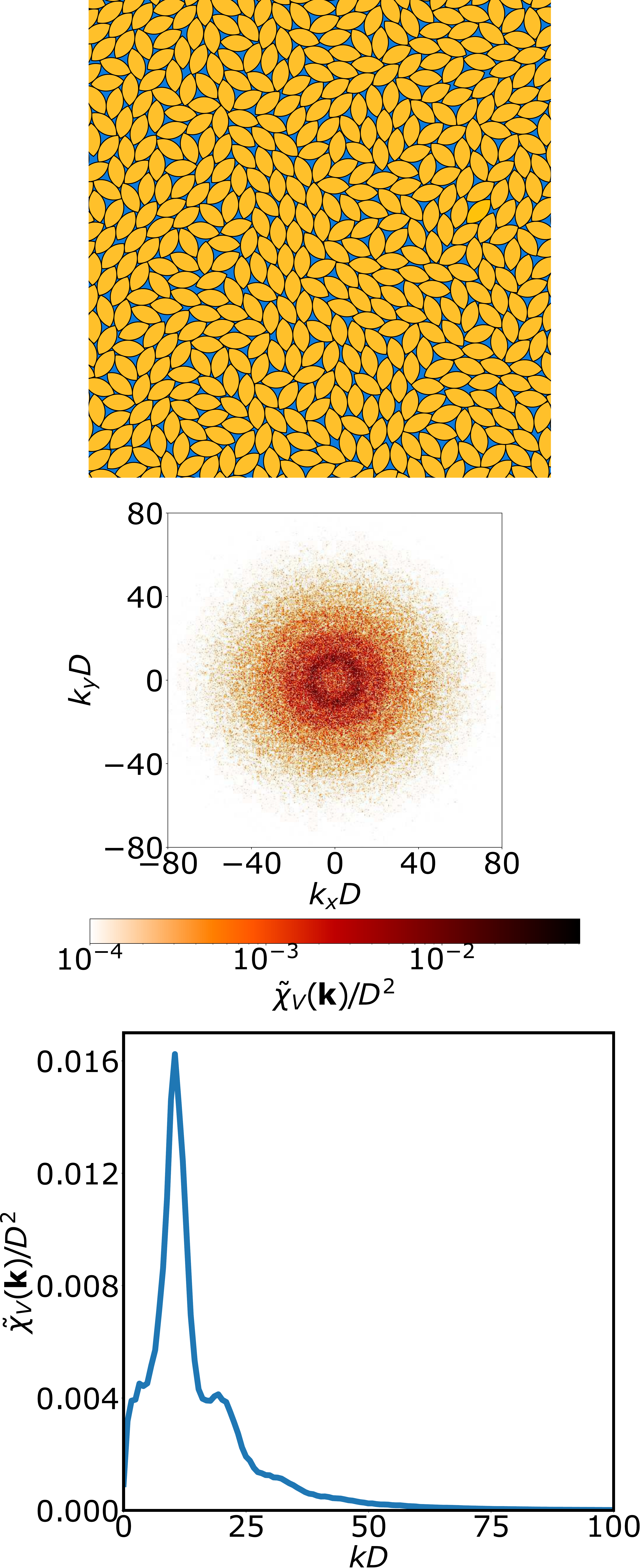}
    \caption{A representative configuration (top), the corresponding two-dimensional (middle), and the angular averaged (bottom) spectral density $\spD{\vect{k}}/D^2$ vs dimensionless wavenumber $kD$ for a rapidly compressed packing of $\alpha=0.5$ lenses, where $D$ is the circumradius of the particle.}
    \label{fig:MRJ?}
\end{figure}

Packings of ICC with $\theta_{ICC}=90^{\circ}$ have short-range translational and 6-fold orientational order that is lost upon angular averaging at all compression rates (see Figure \ref{fig:ICC_Spect}b).
The first peak in the corresponding angular-averaged spectral density corresponds to distance associated with neighboring voids.
As $\theta_{ICC}$ deviates from $90^{\circ}$, the spectral densities become isotropic and exhibit less short-range translational order, which is shown by the disappearance of the peaks in the right panels of Figures \ref{fig:ICC_Spect}a and \ref{fig:ICC_Spect}c.
The broad peaks in these figures are associated with the greater asphericity of the particle shape, resulting in more widely distributed distances between neighboring voids.
The the sharpness of the peak in the right panel of Figure \ref{fig:ICC_Spect}a, compared to that of the right-hand panel of Figure \ref{fig:ICC_Spect}c indicates greater translational order in the $\theta_{ICC}\rightarrow180^{\circ}$ limit than the $\theta_{ICC}\rightarrow0^{\circ}$ limit.
Inspection of the left panel of Figure \ref{fig:ICC_Spect}a shows that at large $\phi$, semicircles do not form circular dimers which could then pack in a triangular lattice, as one may first surmise, suggesting these objects may posses a rich equilibrium phase behavior.

Lens packings with $\alpha > 0.5$ exhibit an increase in short-range 6-fold orientational order, which increases in range as the compression rate decreases, shown by the well-defined peaks in the angular-averaged spectral density in right panel of Figure \ref{fig:Lens_Spect}a.
As $\alpha$ decreases, the degree of short-range 6-fold orientational order decreases, shown by the broadening of the peaks in Figure \ref{fig:Lens_Spect}b.
In these two figures, the first peaks correspond to nearest-neighbor void distances.
For $\alpha$ close to zero, the packings exhibit short-range nematic orientational order (see the center panel of Figure \ref{fig:Lens_Spect}c).
The 6-fold (in Figure \ref{fig:Lens_Spect}a,b) and nematic (in Figure \ref{fig:Lens_Spect}c) orientational order in these packings is lost upon angular averaging.
The corresponding angular-averaged spectral density shows large-scale density fluctuations, which do not decrease in intensity until a length scale associated with the the minor axis of the lens.

It is worth noting that rapidly compressed packings of lenses with $\alpha = 0.5$ have isotropic spectral densities that exhibit little short-range translational order (see Figure \ref{fig:MRJ?}).
The peak in the angular-averaged spectral density is associated with the nearest-neighbor void distances and is broadened due to the orientational disorder in the packing and the anisotropy of the particle shape.
In addition, $\spD{\vect{k}}$ becomes nearly zero ($\approx0.001$), i.e., nearly hyperuniform \cite{Atkinson_Slowdown},  as $k$ approaches 0, implying a suppression of long-wavelength density fluctuations.
This indicates lenses with $\alpha = 0.5$ are a reasonable candidate for a particle shape that readily forms monodisperse disordered jammed packings, an area of current interest \cite{Atkinson_2DMono}.
In three dimensions, the phase diagrams of oblate ellipsoids and lenses are known to closely resemble each other but begin to differ in the hard-sphere limit and for $\phi$ in the crystalline solid range\cite{Lens3D}.
The phase diagram for hard ellipses in ref \citenum{Ellipse_Phase} can thus potentially be used to predict the translational and orientational order in a jammed lens packing by determining if a given $\alpha$ should result in a plastic-to-solid, isotropic-to-solid, or nematic-to-solid phase transition.
In addition, studying the structural changes of these dense packings upon decompression can allow one to qualitatively characterize the equilibrium melting properties of such particles. \cite{TruncTetDense}.

\begin{figure}[t!]
        \centering
        \subfloat[]{\label{fig:Lens_Cont}
        \includegraphics[width = 0.4\textwidth]{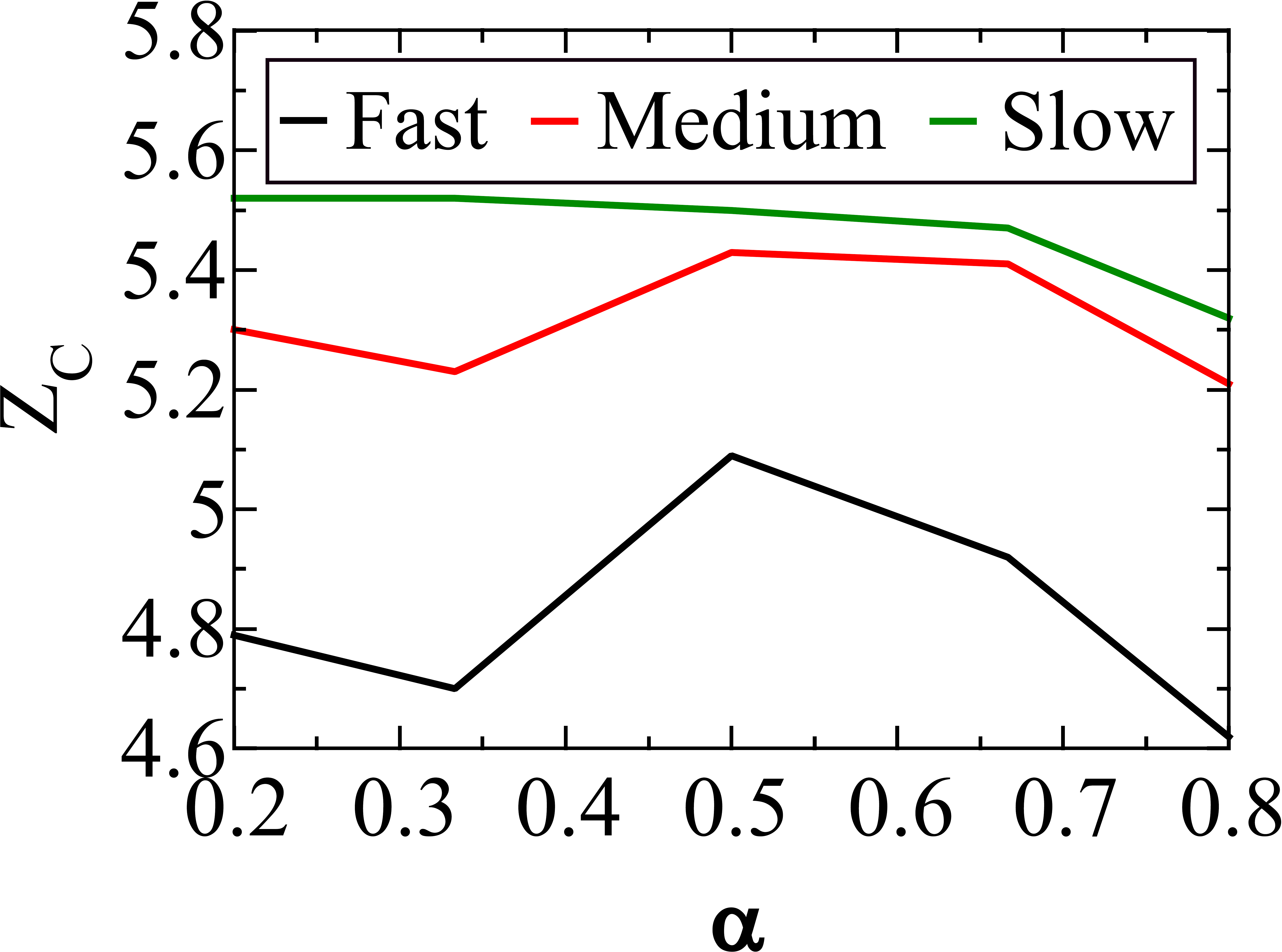}
        }
        
        \subfloat[]{\label{fig:CT_cont}
        \includegraphics[width = 0.4\textwidth]{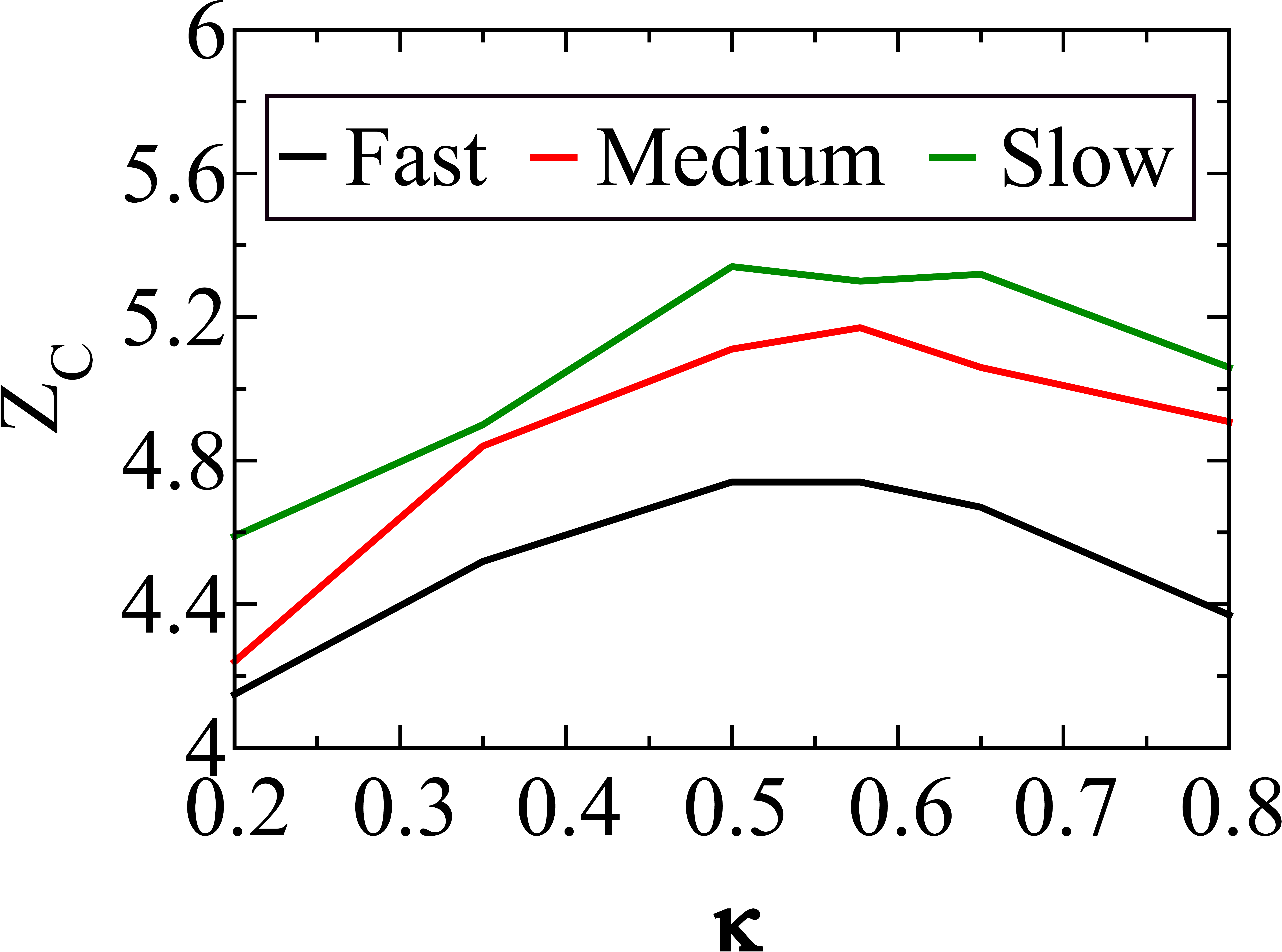}
        }
        \caption{Average number of constraints per particle $Z_C$ in the packings of lenses and CT (ignoring rattlers) as a function of (a) $\alpha$ and (b) $\kappa$.}
        \label{fig:Curved_const}
\end{figure}

Thus, we find that slower compression rates tend to result in longer-range translational and orientational order.
Moreover, we find that packings with large $K$ (see section \ref{sec:Kinetics}) tend to exhibit less short-range translational and orientational order.

\subsection{Contact Statistics}\label{sec:ContactData}

In each of the packings produced by using the ASC scheme, we count the number of each type of contact to determine the contact networks and rattler fraction $\phi_R$.
We then examine how the average number of constraints per particle $Z_C$ and rattler fraction $\phi_R$ change as a function of the particle shape and compression rate.

\begin{figure}[t!]
        \centering
        \subfloat[]{\label{fig:Lens_Rat}
        \includegraphics[width = 0.4\textwidth]{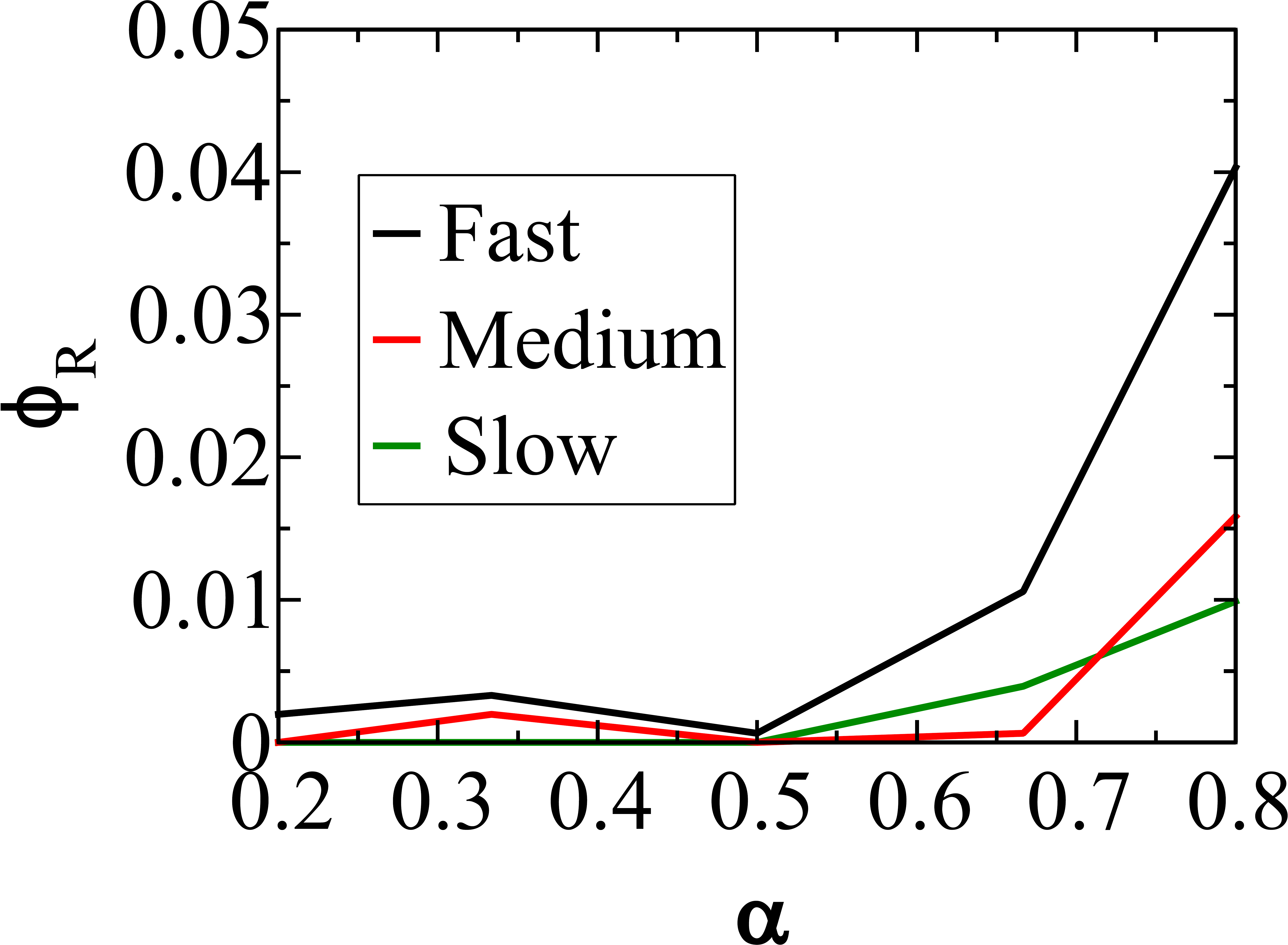}
        }
        
        \subfloat[]{\label{fig:CT_Rat}
        \includegraphics[width = 0.4\textwidth]{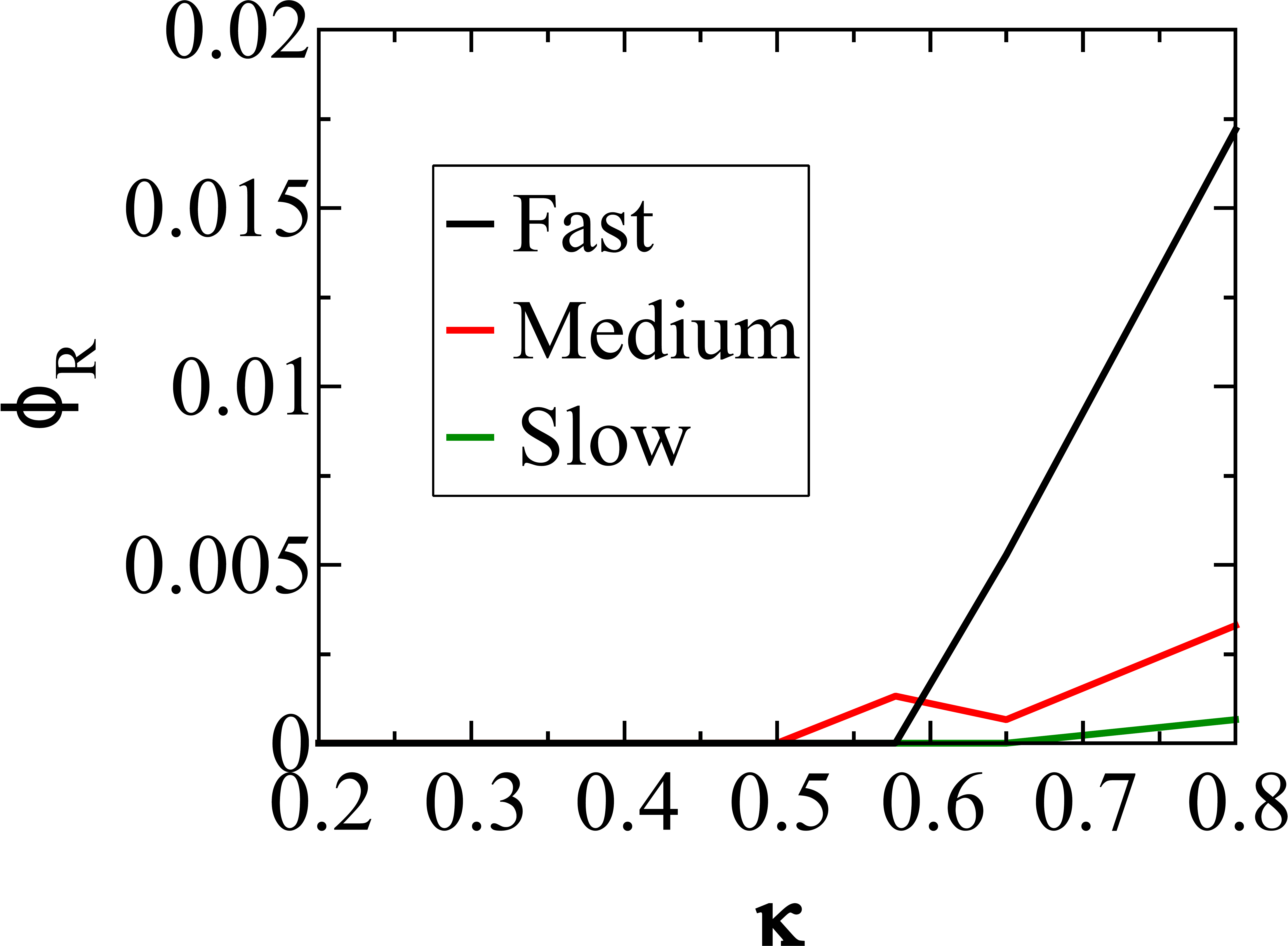}
        }
        \caption{Rattler fraction $\phi_R$ in the packings of lenses and CT as a function of (a) $\alpha$ and (b) $\kappa$.}
        \label{fig:Curved_rat}
\end{figure}

Figure \ref{fig:Curved_const} shows that lens and CT packings are hypostatic, and $Z_C$ decreases as the compression rate increases.
We also observe that $Z_C$ decreases in rapidly compressed packings of lenses with small $\alpha$ and CT with small $\kappa$ because of their larger asphericities $\gamma$ (see Figure \ref{fig:Curved_const}). 
Figure \ref{fig:Curved_rat} demonstrates that $\phi_R$ in these packings tends to increase as the packing is compressed more rapidly and as $\gamma$ decreases. 

\begin{figure}[t!]
        \centering
        \subfloat[]{\label{fig:OST_Const}
        \includegraphics[width = 0.4\textwidth]{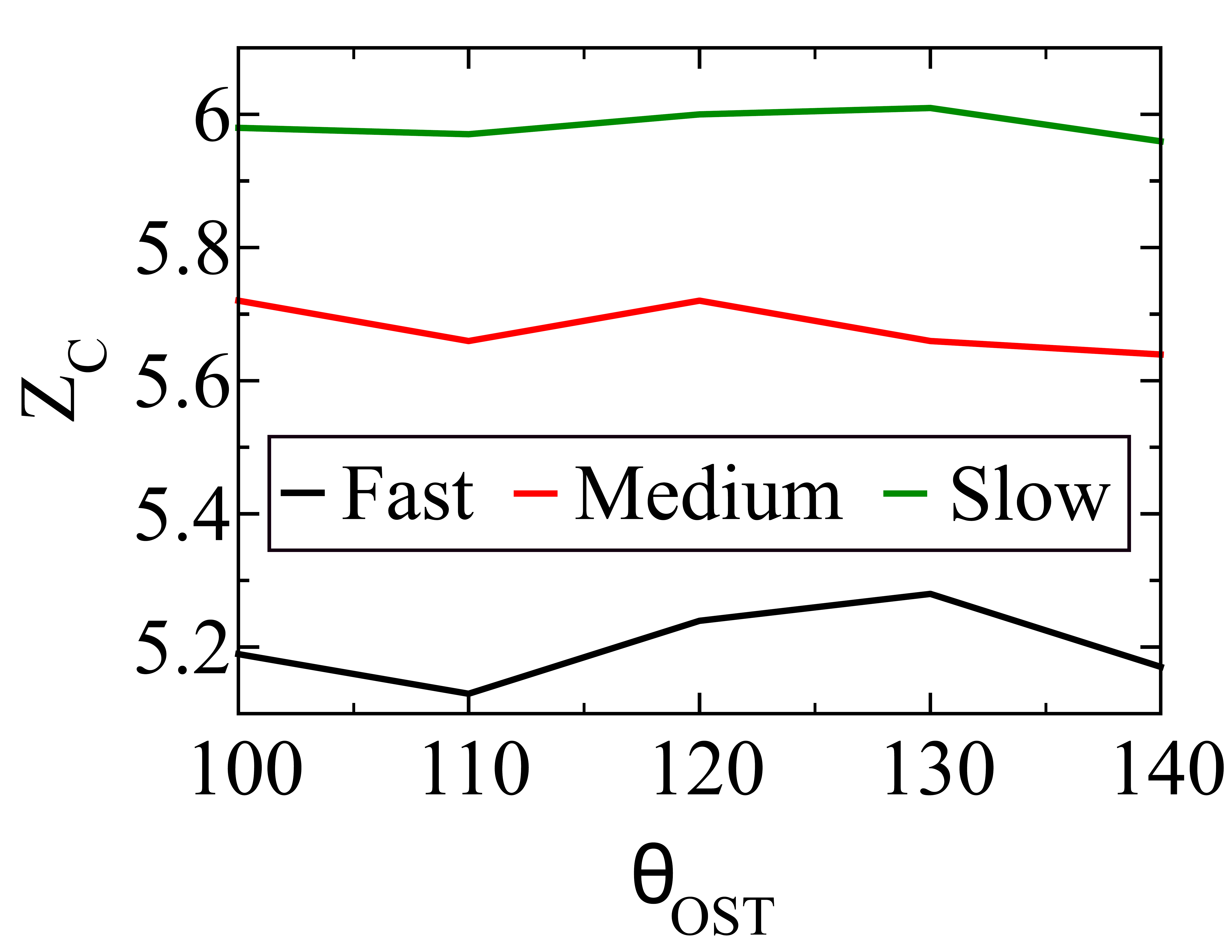}
        }
        
        \subfloat[]{\label{fig:OST_Rat}
        \includegraphics[width = 0.4\textwidth]{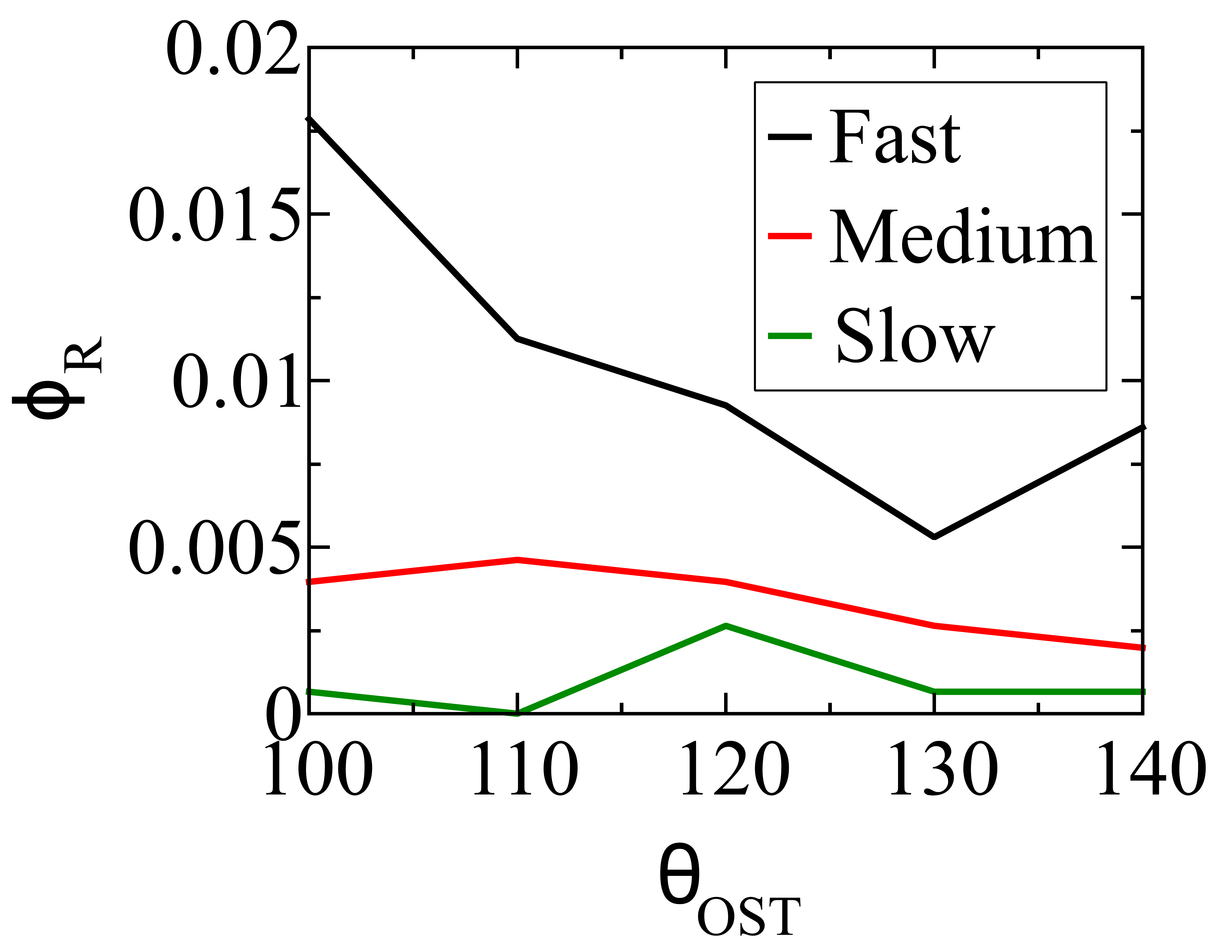}
        }
        \caption{(a) Average number of constraints per particle $Z_C$ (ignoring rattlers) and (b) rattler fraction $\phi_R$ in the OST packings as a function of $\theta_{ICC}$.}
        \label{fig:OST_contres}
\end{figure}

Figure \ref{fig:OST_contres} shows the trends in $Z_C$ and $\phi_R$ as a function of $\theta_{OST}$.
We find that slowly compressed OST packings are nearly isostatic, while all other OST packings are hypostatic.
OST packings have greater $Z_C$ than packings of curved objects, which is attributed to the edge-to-edge contacts only present in packings of faceted particles.
Additionally, we see that $\phi_R$ increases as the packings are compressed more rapidly and as $\gamma$ decreases.

Thus, we find that $Z_C$ tends to increase as the compression rate decreases and $\gamma$ increases. 
Moreover, $\phi_R$ tends to increase as the compression rate increases and as $\gamma$ decreases.

\section{Conclusions}\label{sec:conclusion}

Kinetic effects are ubiquitous in both numerical and experimental hard-particle packing protocols.
We have studied these effects by varying the compression rate used to produce monodisperse packings of convex, noncircular hard-particles in two dimensions.
In particular, we used the ASC scheme to generate dense, effectively jammed packings of rhombus, obtuse scalene triangle, lens, curved triangle, and ice cream cone shaped particles with a wide range of packing fractions and degrees of order.
To characterize the kinetic effects on these packings, we defined the kinetic frustration index $K$ which quantifies the deviation of a packing from its maximum possible packing fraction.
In addition, we characterized the degree of short- and long-range order in the packings using the spectral density.
We also determined the type and number of contacts in the packings to examine the trends in $Z_C$ and $\phi_R$.

Regardless of particle shape, $K$ increases as the compression rate increases.
We found that a higher order of rotational symmetry, smaller $\gamma$, and more curvature all reduce $K$.
Packings with large $K$ tend to have reduced short-range translational order and are more likely to be isotropic.
Additionally, slower compression rates tend to result in packings with a greater degree of short-range translational and orientational order.
We also found that $Z_C$ increases as the $\gamma$ increases and as the compression rate decreases.
Moreover, $\phi_R$ increases as $\gamma$ decreases and as the compression rate increases.

Rapidly compressed lens-shaped particles with $\alpha=0.5$ have spectral densities that indicate short-range translational order, isotropy, and suppressed long-wavelength density fluctuations.
As such, these lenses are a promising candidate for a two-dimensional shape that readily forms monodisperse disordered jammed packings, a current area of interest \cite{Atkinson_2DMono}.
Closer inspection of lenses with $\alpha\sim0.5$ is warranted in future work to determine if this particle shape can readily generate MRJ-like packings.

Our findings on the relationship between $K$ and compression rate may aid in the design of laboratory packing protocols.
It is desirable, however, to extend these results to three-dimensional particle shapes so that they are applicable to a wider variety of physical systems.
Future work in this area should also include the study of kinetic effects in packings in closed containers as well as packings with size and shape polydispersity.
Additionally, it is of interest to determine the equilibrium phase behavior of these families of particle shapes away from the jamming point.
Furthermore, because such hard particle packings can be viewed as two-phase materials, the physical properties (e.g., conductivity and elastic moduli) of these packings should also be studied.
Additional interparticle forces can also be applied to this set of noncircular convex hard particle shapes, e.g., dipolar forces, to design a broader class of packing arrangements (see, e.g. refs \citenum{Hall1} and \citenum{Hall2}.

\begin{figure}[t]
            \centering
            \includegraphics[width=0.4\textwidth]{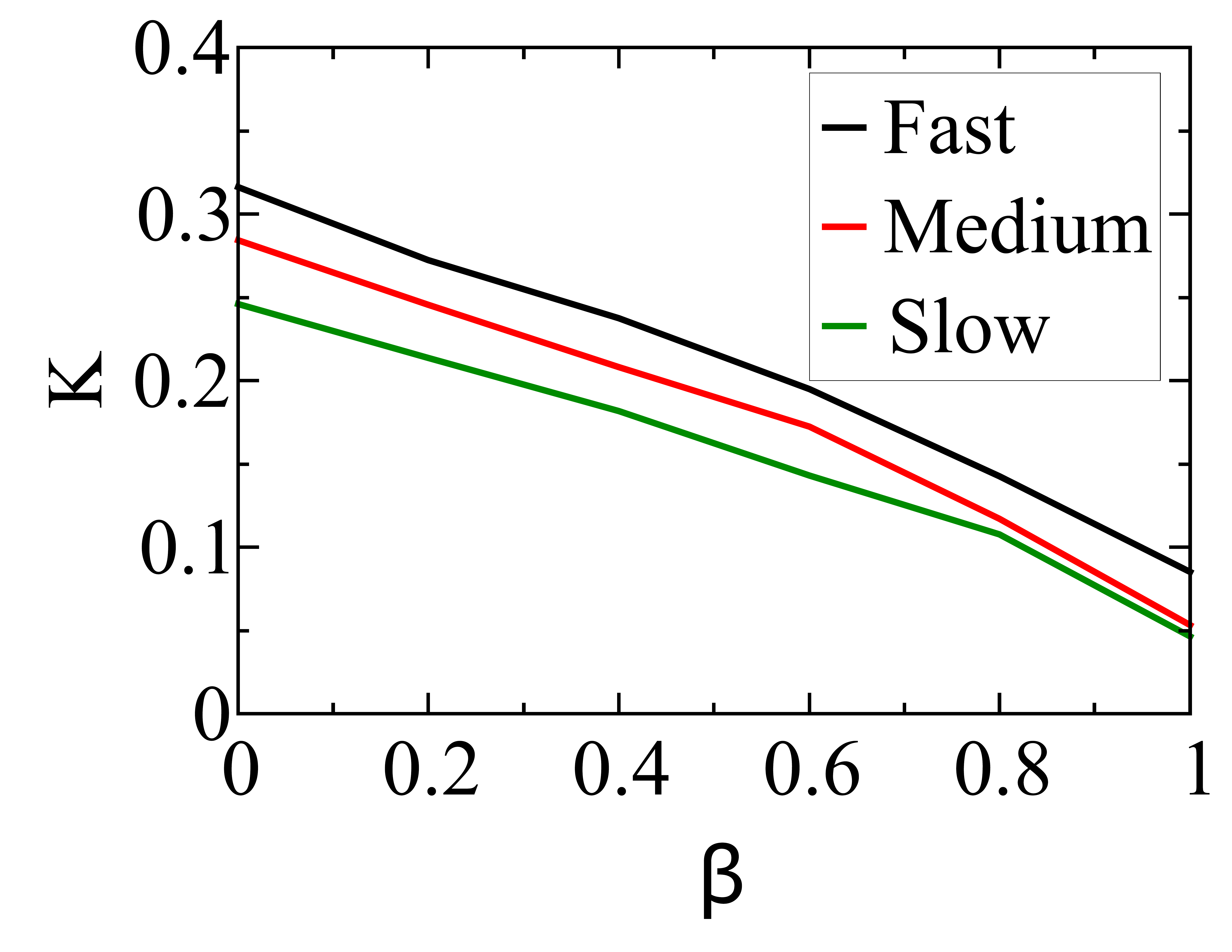}
            \caption{The kinetic frustration index $K$ as a function of thickness $\beta$ for the bowtie shapes.}
            \label{fig:K_bow}
\end{figure}

Obvious nontrivial extensions of the present work include a closer examination of concave two-dimensional shapes and chiral mixtures of particles.
To preliminarily examine the extent to which kinetic frustration is affected by particle concavity, we have measured $K$ for packings of concave bowtie-shaped particles (see Figure \ref{fig:Shape_prot}f) produced using the three compression schedules given in section \ref{sec:ASCScheme}.
Figure \ref{fig:K_bow} shows the behavior of $K$ as a function of thickness $\beta$ for these packings.
We find that a greater difference between the particle volume and the volume of its convex hull results in greater $K$.
It is also noteworthy that bowties with small $\beta$ have the highest values of $K$ observed herein.
This indicates concave particle shapes tend to more easily generate packings with large $K$.
Thus, such particle shapes may allow for the generation of effectively jammed packings with small packing fractions, which warrants further examination.

Moreover, becuase of the wide range of enantioselectivities in the synthesis of chiral molecules \cite{ojima_chiral}, it is of interest to determine the effect of enantioenrichment on packing kinetics.
Here, we preliminarily consider dense packings of enantiopure (a system containing only one type of mirror image) and racemic (a system containing an equal amount of both mirror images) packings of OST with $\theta_{OST}=100^{\circ}$.
Table \ref{tab:chiral} compares the values of the kinetic frustration index $K$ for these two types of packings.
We find that packings of racemic mixtures of OSTs have a higher $K$ than the analogous enantiopure packings.
Further studies of the effects of chirality on kinetic frustration include the study of $K$ as a function of enantioenrichment as well as the examination of other chiral particle shapes.

\begin{table}[t]
\centering
\caption{$K$ Values for OSTs with $\theta_{OST}=100$ for All Three Compression Schedules}
\begin{tabular}{|c|c|c|}
\hline
                     & \multicolumn{2}{c|}{$K$}                \\ \hline
compression rate     & racemic                   & enantiopure \\ \hline
fast                 & 0.140612                  & 0.127592    \\ \hline
medium               & 0.110041                  & 0.096288    \\ \hline
slow                 & 0.096090                  & 0.084675    \\ \hline
\end{tabular}
\label{tab:chiral}
\end{table}

\begin{acknowledgement}

The authors are grateful to Duyu Chen, Murray Skolnick, and Steven Atkinson for assistance in implementing the ASC code and to Michael A. Klatt and Amy Secunda for fruitful discussions. This work was supported by the by the National Science Foundation (NSF) under Grant DMR-1714722. The authors declare no competing financial interest.

\end{acknowledgement}



\providecommand{\noopsort}[1]{}\providecommand{\singleletter}[1]{#1}%
\providecommand{\latin}[1]{#1}
\makeatletter
\providecommand{\doi}
  {\begingroup\let\do\@makeother\dospecials
  \catcode`\{=1 \catcode`\}=2 \doi@aux}
\providecommand{\doi@aux}[1]{\endgroup\texttt{#1}}
\makeatother
\providecommand*\mcitethebibliography{\thebibliography}
\csname @ifundefined\endcsname{endmcitethebibliography}
  {\let\endmcitethebibliography\endthebibliography}{}


\begin{tocentry}

\includegraphics[width=3.25in]{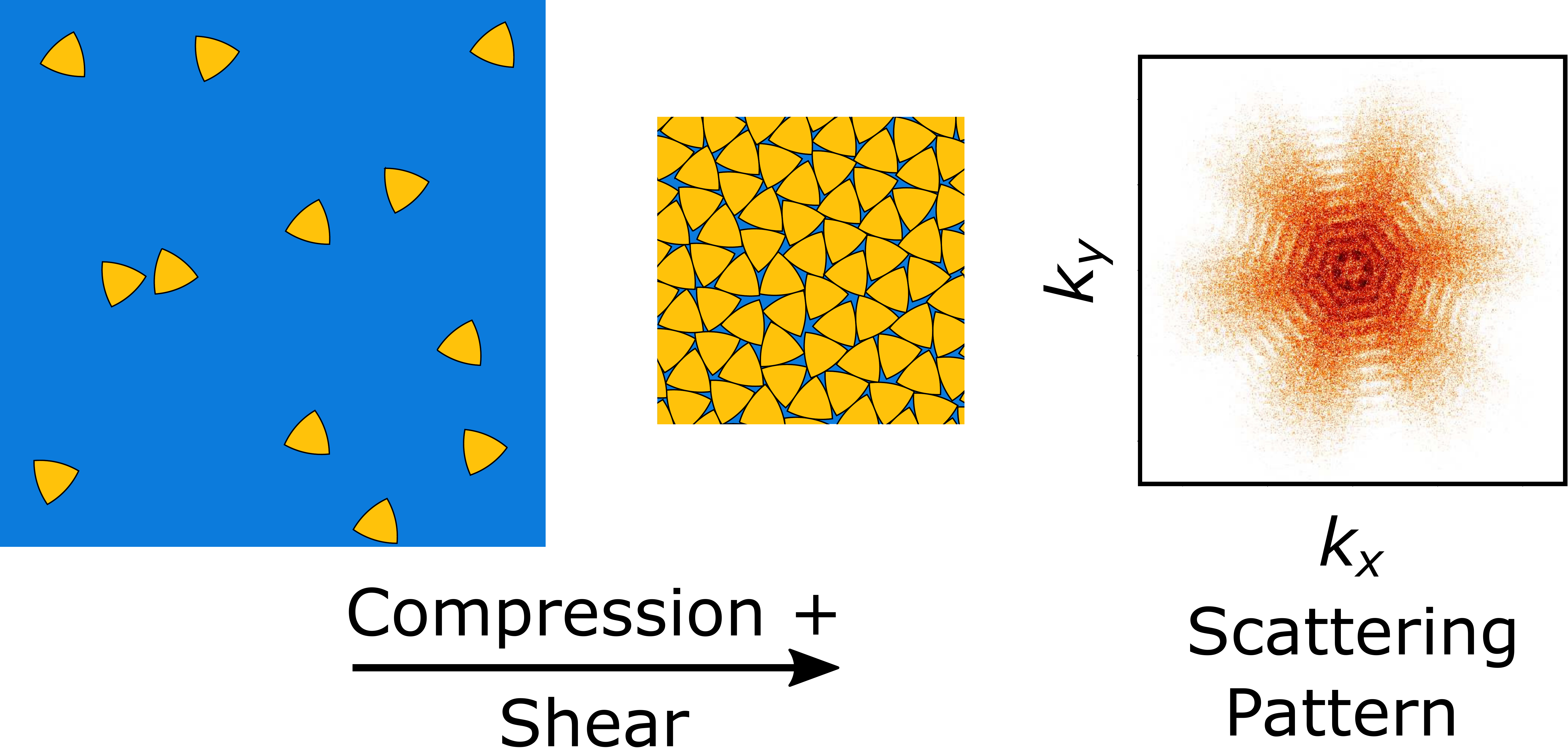}

\end{tocentry}

\end{document}